\documentclass[10pt,oneside,onecolumn,a4paper,draftclsnofoot]{IEEEtran}
\usepackage{epsfig,bbm}
\usepackage{amsmath}
\usepackage{amsfonts}
\usepackage{gensymb}
\usepackage{graphicx}
\usepackage{xcolor}
\usepackage[ruled,linesnumbered]{algorithm2e}
\usepackage{multirow}
\def\comment#1{}

\newcommand{\A}{{\bf A}}
\newcommand{\W}{{\bf W}}
\newcommand{\X}{{\bf X}}

\newcommand{\R}{{\bf R}}
\newcommand{\y}{{\bf y}}
\newcommand{\z}{{\bf z}}
\newcommand{\x}{{\bf x}}
\newcommand{\s}{{\bf s}}
\newcommand{\g}{{\bf g}}
\newcommand{\h}{{\bf h}}
\newcommand{\I}{{\bf I}}

\newcommand{\tr}{{\tt tr}}
\renewcommand{\a}{{\bf a}}
\newcommand{\C}{{\bf C}}
\newcommand{\Cx}{{\bf C}_{\bf x}}
\newcommand{\wCx}{\widehat{\bf C}_{\bf x}}
\newcommand{\Cz}{{\bf C}_{\bf z}}
\newcommand{\Czz}{{\bf C}_\mathbbm{z}}
\newcommand{\wCz}{\widehat{\bf C}_{\bf z}}

\newcommand {\bxi} {\boldsymbol{\xi}}

\begin{document}

\title{Gradient Algorithms for Complex Non-Gaussian Independent 
	Component/Vector Extraction, Question of Convergence}

\author{{\bf Zbyn\v{e}k Koldovsk\'{y}$^1$ and Petr 
		Tichavsk\'y$^2$}
	\vspace{0.1in} \\
	$^1$Acoustic Signal Analysis and Processing Group, Faculty of 
	Mechatronics, 
	Informatics, and Interdisciplinary\\ Studies,
	Technical University of Liberec, Studentsk\'a 2, 461 17
	Liberec, Czech Republic. \\E-mail:
	zbynek.koldovsky@tul.cz, fax:+420-485-353112, tel:+420-485-353534\\
	%$^2$Conexant System, 1901 Main Street, Irvine, CA (USA)
	$^2$The Czech Academy of Sciences, Institute of Information Theory and 
	Automation,\\ Pod
	vod\'{a}renskou v\v{e}\v{z}\'{\i} 4, P.O.~Box 18, 182 08 Praha 8,
	Czech Republic. E-mail: tichavsk@utia.cas.cz,
	fax:+420-2-868-90300, tel. +420-2-66052292
}

%\vspace{-1cm}

\maketitle

\footnotetext{Part of this paper was presented at the 25th European Signal 
	Processing Conference (EUSIPCO 2017) in Kos, Greece. This work was 
	supported by 
	The Czech Science Foundation through Project No.~17-00902S.}

{\color{black}
	\begin{abstract}
We revise the problem of extracting one independent component from 
an 
instantaneous linear mixture of signals.
The mixing matrix is parameterized by two vectors, one column of 
the mixing matrix and one row of the de-mixing matrix. %, like in 
%the Internal Linear Combination (ILC) method
%for extraction of a single component, but the difference is that 
%no part of the mixing matrix is known a priori.
%Compared to Independent Component 
%Analysis, a novel parameterization of the %mixing model is used. 
The separation is based on the non-Gaussianity of the source of 
interest, while the other 
background signals are assumed to be Gaussian. 
Three gradient-based estimation algorithms are derived using the 
maximum 
likelihood principle and are compared with the Natural Gradient 
algorithm for Independent Component Analysis and with One-unit 
FastICA based on negentropy maximization. The ideas and algorithms 
are also generalized for 
the extraction of a vector component when 
the extraction proceeds jointly from a set of instantaneous 
mixtures.
Throughout the paper, we address the problem of the size of the 
region of convergence for 
which the algorithms guarantee the extraction of the desired 
source. We show how that size is influenced by the ratio of powers 
of the sources within the mixture. Simulations confirm this 
observation where several algorithms are compared. They show 
various convergence behaviour in a scenario where the source of 
interest is dominant or weak. Here, our proposed modifications of 
the gradient methods taking into account the dominance/weakness of 
the source show improved global convergence property. 
	\end{abstract}
}

{\keywords Blind Source Separation, Blind 
	Source Extraction, Independent Component Analysis, Independent Vector 
	Analysis}

\section{Introduction}
\subsection{Independent Component Analysis}
Independent Component Analysis has been a popular method studied for Blind 
Source Separation (BSS)
since the 1990s 
\cite{lee1998,hyvarinen2001,cichocki2002,comon2010handbook}. 
In the basic ICA model, 
signals observed on $d$ sensors are assumed to be linear mixtures of $d$ 
``original'' signals, which are mutually independent in the statistical 
sense. 
The mixing model is given by
\begin{equation}\label{ICA}
\x(n) = \A {\bf u}(n),  
\end{equation} 
where $\x(n)$ is a $d\times 1$ vector of the mixed signals; $\A$ is a 
$d\times 
d$ nonsingular mixing matrix; ${\bf u}(n)$ is a $d\times 1$ vector of the 
original signals; and $n$ denotes the sample index. In the 
non-Gaussianity-based ICA, the $j$th original signal $u_j(n)$ (the $j$th 
element of ${\bf u}(n)$) is modeled as an independently and identically 
distributed (i.i.d.) sequence of random variables with the probability 
density 
function (pdf) $p_j(\cdot)$. The goal is to estimate $\A^{-1}$ using 
$\x(n)$, 
$n=1,\dots,N$, through finding a square de-mixing matrix ${\bf W}$ such 
that 
${\bf y}(n)={\bf Wx}(n)$ are independent or as close to independent as 
possible. In the 
discussion that follows, we will omit the sample index $n$ for the sake of 
brevity, except where it is 
required.

While our focus in this paper is on complex-valued signals and parameters, 
our 
conclusions are valid for the real-valued case as well.

\subsection{Blind Extraction of One Independent Source}
This work addresses the problem of extraction (separation) of one 
independent component, which is often sufficient in applications such as 
speaker source enhancement, passive radar and sonar, or in 
biomedical signal processing. The complete decomposition performed by ICA 
can 
be computationally very demanding and superfluous. This is 
especially remarkable when there is a large number of sensors (say, 10 or 
more), 
or when a large number of mixtures (say, 128 or more) are separated in 
parallel, as in the Frequency-Domain ICA (FD-ICA) \cite{smaragdis1998}. The 
idea of extracting only one source can also be applied in joint BSS 
\cite{li2009,lahat2014,li2011}, especially in Independent Vector Analysis 
(IVA) \cite{kim2007}. Here, the ``source'' is represented by a vector of 
separated components from the mixtures that are mutually dependent (but 
independent of the other vector components).

{\color{black} %One important application of separation of one independent 
%source is separation of the cosmic microwave background from the Wilkinson 
%Microwave Anisotropy Probe (WMAP) data or from the more recent Planck 
%mission data, see \cite{remazeilles2011,cardoso2008} and the references 
%therein. The mathematical model is nearly the same as in this paper, 
%however, there is one important difference that the steering vector of the 
%signal-of-interest, sometimes called ``mixing vector", is assumed to be 
%known. The methods are known under the name Internal Linear Combination 
%(ILC). %Their goal is to estimate the component of the interest under the 
%%assumption of the partial knowledge of the mixing matrix, and assumption 
%%of 
%%statistical independence of the SOI and the background.
	
	%In this paper, the mixing vector is subject of blind estimation, which 
	%is harder task to achieve than the goal of ILC. Still we assume that, 
	%indeed, some partial knowledge about the SOI must be available.
	BSS involves the permutation ambiguity 
	\cite{hyvarinen2001,sawada2004icakyoto}, and therefore some partial 
	knowledge about the SOI must be available to determine which 
	independent component is the one of our interest.}
For example, the a~priori knowledge could be an expected direction of 
arrival (DOA) 
of the source, location of the source within a confined area 
\cite{koldovsky2013}, a property such 
as dominance within an angular range \cite{sawada2005icassp} or temporal 
structure \cite{javidi2010}, 
and so forth. {\color{black} Throughout this paper, we will assume that 
such 
	knowledge is available in the form of an 
	initial value of a (de)mixing parameter.} The wanted signal will be 
	referred to 
as {\em source of interest (SOI)} while the rest of the mixture will be 
referred to as {\em background}.

{\color{black} The theoretical part of this paper will be constrained to 
the 
	determined case, which means that the background is assumed to be a 
	mixture of 
	$d-1$ latent variables, or, in other words, the whole mixture obeys the 
	determined mixing as in \eqref{ICA}.} This assumption need not be 
	overly 
restrictive when only 
one source should be extracted. Indeed, when the mixture $\x$ consists of 
more 
than $d$ sources (underdetermined case), algorithms based on the determined 
model can still be used provided that they are sufficiently robust against 
mismodeling and noise.  {\color{black} However, these issues are beyond the 
scope of this paper.}

\subsection{State-of-the-Art}\label{sec:stateoftheart}

The blind separation of one particular non-Gaussian source has already been 
studied 
in several contexts and some authors refer to it as Blind Signal Extraction 
(BSE) \cite{amari1998b,cruces2004,javidi2010}. Projection pursuit 
\cite{huber1985}, a technique 
used for exploratory data analysis, aims at finding ``interesting'' 
projections, 
including 1-D signals. This ``interestingness'' is defined through various 
measures 
reflecting the distance of the projected signal's pdf from the Gaussian 
distribution \cite{hyvarinen2001,girolami1997}. Various criteria of the 
interestingness were also derived in other contexts. 
For example, kurtosis 
appears in methods for blind adaptive beamforming or as a higher-order 
cumulant-based contrast; see, e.g., \cite{shalvi1993,shynk1996,moreau1996}.

This framework was unified under ICA based on 
information theory \cite{cover}. Namely, the independence of signals can 
be measured using mutual information, which is the Kullback-Leibler 
divergence 
between the joint density of signals and the product of their marginal 
densities. The signals are independent if and only if that mutual 
information 
equals 
zero. {\color{black} Provided that the elements of $\y(n)$ are not 
correlated}, the mutual 
information of ${\bf y}(n)$ is equal to the sum of entropies of 
$y_1(n),\dots,y_d(n)$, up to a constant. Hence, it follows that an 
independent 
component can be sought through minimizing the entropy of the separated 
signal 
under the constraint ${\rm E}[|{\bf w}^H\x|^2]=1$; here ${\rm E}[\cdot]$ 
stands 
for 
the expectation operator; and ${\bf w}^H$ is a de-mixing vector (a row of 
the 
de-mixing matrix $\W$). 
The fact that entropy is a measure of non-Gaussianity 
reveals the connection between the ICA-based
separation of one signal and the contrast-based BSE techniques 
\cite{hyvarinen2001,cardoso1998}.

In fact, many ICA methods apply $d$ BSE estimators sequentially 
\cite{delfosse1995} or in parallel \cite{erdogan2009} to find all 
independent 
components in the mixture. The orthogonal constraint, which requires that 
the 
sample correlation between separated signals is equal to zero, is imposed 
on 
the BSE outputs in order to prevent the algorithms from finding any 
components 
twice. For 
example, the well-known FastICA algorithm has three basic variants: 
One-Unit 
FastICA is a BSE method optimizing the component's non-Gaussianity 
\cite{hyvarinen1997}; Deflation FastICA applies the 
one-unit version sequentially \cite{hyvarinen1999}; Symmetric FastICA runs 
$d$ 
one-unit algorithms simultaneously \cite{hyvarinen1999,wei2015}. 

The separation accuracy of the above methods is known to be limited 
\cite{tichavsky2006}. 
One-unit FastICA exploits only the non-Gaussianity of SOI and does not use 
the 
non-Gaussianity of the background 
\cite{hyvarinen1997b}. The accuracy levels of deflation and symmetric 
FastICA 
are limited due to the orthogonal constraint \cite{cardoso1994}. While the 
latter limitation can be overcome, as shown, e.g., in 
\cite{koldovsky2006}, the limited accuracy of the one-unit approach poses 
an 
open problem, unless the BSE is done through the complete ICA. By comparing 
the 
performance analyses from 
\cite{hyvarinen1997b,pham2006,tichavsky2006} and the Cram\'er-Rao bound for 
ICA 
\cite{koldovsky2006}, it follows that one-unit methods can approach the 
optimum 
performance only when the background is Gaussian, but not otherwise.

\subsection{Contribution}
{\color{black} In this paper, we revisit the BSE problem by considering it 
	explicitly as the 
	goal to extract one component from the instantaneous mixture that is as 
	close 
	to being 
	independent of the background as possible; we refer to this approach as 
	{\em 
		Independent Component Extraction (ICE)}. A re-parameterization of 
		the mixing 
	model is introduced, in which the number of 
	parameters is minimal for the BSE problem (the mixing and the 
	separating vector related to the SOI).  Then, a statistical model is 
	adopted from ICA where the background is assumed to be jointly 
	Gaussian. The 
	classical maximum likelihood estimation of the mixing parameters is 
	considered, 
	by which simplistic gradient-based estimation algorithms are 
	derived\footnote{{\color{black} In this paper, an algorithm estimating 
	the 
			separating vector is introduced, compared 
			to \cite{koldovsky2017b}, where only the variant for the mixing 
			vector 
			estimation is described.}}.  The ICE approach provides a deeper 
			insight into 
	the BSE problem. In particular, it points to the role of the orthogonal 
	constraint and to the fact that the constraint is inherently applied 
	within One-Unit FastICA. The approach also reveals the role of the 
	model of background's pdf. 
	
	{\color{black} 
		It is worth pointing out here that a similar mixing and statistical 
		model have been considered in methods that were designed for Cosmic 
		Microwave Background extraction from the Wilkinson Microwave 
		Anisotropy Probe (WMAP) data or from the more recent Planck Mission 
		data \cite{cardoso2017}. However, there is one important difference 
		that the mixing vector related to the SOI is assumed to be known. 
		The methods are known under the name Internal Linear Combination 
		(ILC) \cite{dick2010,remazeilles2011}; see also \cite{cardoso2008}.}
	
	For the practical output of this paper, we focus on the ability of BSE 
	algorithms to ensure that the desired SOI is being extracted, not a 
	different 
	source\footnote{\color{black}  We do not focus on the algorithms' 
	accuracy as 
		this is already a well-studied problem. The accuracy of BSE methods 
		is 
		fundamentally limited by the Cram\'er-Rao bound, which is 
		asymptotically 
		attainable, e.g., by One-Unit FastICA \cite{tichavsky2006}; cf. the 
		last 
		paragraph of 
		Section~\ref{sec:stateoftheart}; see also \cite{kautsky2017}.}. 
		This is a 
	crucial aspect in BSE, which has been little studied previously. 
	When the extraction of the SOI is not guaranteed, it is necessary to 
	extract 
	all sources and to find the desired one afterwards, in which way the 
	advantage of doing only one BSE task is lost. The other motivation is 
	that the permutation ambiguity can impair on-line separation. For 
	example, a sudden change of the region of convergence (ROC) due to 
	dynamic signals and/or mixing conditions can cause that the current 
	mixing vector estimate occurs within the ROC of a different source. 
	Then, the given algorithm performs several ``diverging'' steps during 
	which the separated sources are being permuted; the separation is poor 
	in the meanwhile.
	%In fact, the tasks to find the desired source among the separated 
	%sources (the solution of the permutation problem) or to control the 
	%convergence 
	%of the given BSE algorithm are similar.
	%important when extracting only one SOI. The desired property of a BSE 
	%method 
	%is that it extracts the SOI with as little sensitivity to the initial 
	%bias as 
	%possible. 
	Therefore, the size of the ROC to the SOI is studied. We 
	point to the fact that the ROC is algorithm-dependent and is highly 
	influenced by the so-called Scales 
	Ratio (SR), which is the ratio of the powers of the SOI and the 
	background. The experiments show that the ROC can depend on whether the 
	optimization proceeds in the mixing or de-mixing parameters. Based on 
	this, we propose novel variants of the gradient algorithms where the 
	optimization parameters are selected automatically.

	Next, the ideas are generalized to the extraction of a vector 
	component, 
	so-called {\em Independent Vector Extraction (IVE)}. Here, the 
	problem defines several instantaneous mixtures to be treated 
	simultaneously 
	using joint statistical models. The goal is to extract one independent 
	component per mixture where the extracted components should be as 
	dependent as 
	possible. IVE is an extension of ICE similar to that of ICA to IVA 
	\cite{lee2007iva,adali2014}. A gradient algorithm with the automatic 
	selection 
	of the optimization parameters is derived, similarly to that for ICE. 
	The 
	experiments show that the convergence of the proposed IVE algorithm is 
	superior 
	to that for ICE because improved convergence within several mixtures 
	has 
	positive influence on the convergence within the other mixtures; the 
	effect of 
	the automatic selection is thus multiplied.}

The rest of this paper is organized as follows. Section~II introduces  
algebraic and statistical models for ICE. Section~III is devoted to 
gradient-based ICE algorithms. The ideas and algorithms are generalized to 
the 
extraction of vector components in Section~IV. Section~V presents results 
of 
simulations, and Section VI concludes the paper.

\section{Independent Component Extraction}
\subsubsection*{Nomenclature} The following notation will be used 
throughout 
the article. Plain letters 
denote scalars, bold lower-case letters denote vectors, and bold capital 
letters denote 
matrices. The Matlab conventions for matrix/vector concatenation and 
indexing 
will be used, e.g., $[1;\,{\bf g}]=[1,\, {\bf g}^T]^T$, and $(\A)_{j,:}$ is 
the 
$j$th row of $\A$. Next, $\g^T$, $\overline{\g}$, and $\g^H$ denote the 
transpose, the complex conjugate value and the conjugate transpose of $\g$, 
respectively. 
Symbolic 
scalar and vector random variables will be denoted by lower-case letters, 
e.g., 
${s}$ and ${\bf x}$, while the quantities collecting their $N$ samples will 
be 
denoted by bold (capital) letters, e.g., ${\bf s}$ and ${\bf X}$. Estimated 
values of signals will be denoted by hats, e.g., $\widehat{\bf s}$. For 
simplicity, the hat will be omitted in the case of estimated values of 
parameters, 
e.g., 
${\bf w}$, unless it is necessary to distinguish between its estimated and 
true 
values.

\subsection{Mixing Model Parameterization}\label{sec:mixing}
Without any loss of generality, let the SOI be $s=u_1$ and $\a$ be 
the first column of $\A$, so it can be partitioned as $\A=[\a,\,\A_2]$. 
Then, 
$\x$ can be written in the form
\begin{equation}\label{modelsinglesource}
\x=\a s + \y,
\end{equation}
where ${\bf y}=\A_2{\bf u}_2$ and ${\bf u}_2=[u_2,\dots,u_d]^T$. The
single-target description \eqref{modelsinglesource} has been widely studied 
in 
array processing literature \cite{vantrees2002}. Here, the fact that 
$\y=\A_2{\bf u}_2$ means that we restrain our considerations to the 
determined 
scenario (the mixture consists of the same number of sources as that of 
the sensors).

Let the new mixing matrix for ICE and its inverse matrix 
be denoted by $\A_{\rm ICE}$ and $\W_{\rm ICE}$, respectively. In ICE, 
the identification of $\A_2$ or the decomposition of ${\bf y}$ into 
independent 
signals is {\em not} the goal. Therefore, the structure of the mixing 
matrix is 
$\A_{\rm ICE}=[{\bf a},\, {\bf Q}]$ where ${\bf Q}$ is, for now, arbitrary. 

Then, \eqref{modelsinglesource} can be written as 
\begin{equation}\label{mixingICE}
\x = \A_{\rm ICE}{\bf v},
\end{equation}
where ${\bf v}=[s;{\bf z}]$, and ${\bf y}={\bf Q}{\bf z}$. It holds that 
${\bf 
	z}$ spans the same subspace as that spanned by ${\bf u}_2$. 

To complete the mixing model definition, we look at the inverse matrix 
$\W_{\rm 
	ICE}=\A_{\rm ICE}^{-1}$. Let ${\bf a}$ and $\W_{\rm ICE}$ be 
	partitioned, 
respectively, as 
\begin{equation}
\a=\begin{pmatrix}
\gamma\\\g
\end{pmatrix}
\end{equation} 
and 
\begin{equation}
\W_{\rm ICE}=\begin{pmatrix}
{\bf w}^H\\ {\bf B}
\end{pmatrix}.
\end{equation} 
${\bf B}$ is required to be orthogonal to ${\bf a}$, i.e., ${\bf Ba}={\bf 
	0}$, which ensures that the 
signals separated by the lower part of $\W_{\rm ICE}$, namely, by ${\bf 
B}\x$, 
do 
not contain any contribution of $s$. A useful selection is 
\begin{equation}
{\bf B}=\begin{pmatrix}
{\bf g} & -\gamma{\bf I}_{d-1}
\end{pmatrix},
\end{equation}
where ${\bf I}_d$ denotes the $d\times d$ identity matrix. 
Let ${\bf w}$ be partitioned as
\begin{equation}
{\bf w}=\begin{pmatrix}
\beta\\\h
\end{pmatrix}.
\end{equation} 
The de-mixing matrix then has the structure
\begin{equation}\label{demixing}
\W_{\rm ICE}=
\begin{pmatrix}
{\bf w}^H\\
{\bf B}
\end{pmatrix}=
\begin{pmatrix}
\overline{\beta} & {\bf h}^H\\
\g & -\gamma{\bf I}_{d-1}
\end{pmatrix},
\end{equation}
and from $\A_{\rm ICE}^{-1}={\bf W}_{\rm ICE}$ it follows that
\begin{equation}\label{mixing}
\A_{\rm ICE}=\begin{pmatrix}{\bf a} & {\bf Q}\end{pmatrix}=
\begin{pmatrix}
\gamma & {\bf h}^H\\
\g & \frac{1}{\gamma}\left({\bf g}{\bf h}^H-{\bf I}_{d-1}\right)
\end{pmatrix},
\end{equation}
where $\beta$ and $\gamma$ are linked through 
\begin{equation}\label{betagamma}
\overline{\beta}\gamma=1-{\bf h}^H{\bf g}.
\end{equation} 
The latter equation can also be written in the form ${\bf w}^H{\bf a}=1$, 
which 
is known as the {\em distortionless response} constraint; see page 515 in  
\cite{vantrees2002}. {\color{black} The parameterization of the mixing and 
de-mixing matrix is similar to the one used in ILC 
\cite{remazeilles2011,cardoso2008}.}

It is worth mentioning here that ICE and Multidimensional ICA 
\cite{cardoso1998b} are similar to each other; the latter is also known as 
Independent Subspace Analysis (ISA) 
\cite{hyvarinen2006}. In ISA, the goal is to separate subspaces of 
components 
that are mutually independent while components inside of the subspaces can 
be 
dependent. The goal to separate one independent component thus could be 
formulated as a special case of ISA where ${\bf u}$ is divided into two 
subspaces of dimensions $1$ and $d-1$, respectively. What makes ICE 
different 
is 
that the separation of the background subspace from the SOI is not as 
ambiguous 
as in ISA, which is ensured by the structure of the de-mixing matrix.

The free variables in the ICE mixing model are the elements of ${\bf g}$ 
and 
${\bf h}$, and one of the parameters $\beta$ or $\gamma$. In total, there 
are 
$2d-1$ free (real or complex) parameters. 
The role of $\a$, as follows from \eqref{modelsinglesource}, is the {\em 
mixing 
	vector} related to $s$, which is in beamforming literature also 
	sometimes 
referred to 
as the steering vector.
Next, ${\bf w}$ is the {\em separating vector} as $s={\bf w}^H\x$. For the 
background signal $\z$, it holds that
\begin{equation}\label{background}
\z={\bf B}\x={\bf B}\y={\bf B}\A_2{\bf u}_2.
\end{equation}
Note that $\A_2$ is not identified in the model, so the relationship 
between 
$\z$ 
and ${\bf u}_2$ remains unknown after performing ICE. The components of 
$\z$ 
are 
independent after the extraction only when $d=2$ or, for $d>2$, in very 
special 
cases that ${\bf B}\A_2=\boldsymbol{\Lambda}{\bf P}$; 
$\boldsymbol{\Lambda}$ denotes a diagonal (scaling) matrix with non-zero 
elements on the main diagonal, and ${\bf P}$ denotes a permutation matrix.

\subsection{Indeterminacies}\label{sec:indeterminacies}
Similarly to ICA, the scales of $s$ and of ${\bf a}$ are ambiguous in the 
sense 
that, in \eqref{modelsinglesource} they can be replaced, respectively, by 
$\alpha s$ and $\alpha^{-1}{\bf a}$ where $\alpha\neq 0$. The scaling 
ambiguity 
can be avoided by fixing $\beta$ or $\gamma$. A specific case occurs when 
$\gamma=1$ \cite{koldovsky2017b}, since then 
$s$ corresponds to the so-called spatial image of the SOI on the first 
sensor 
\cite{matsuoka2001,duong2010under,koldovsky2017}. 
This can be useful for modeling the pdf of $s$, as the physical meaning of 
that 
scale is often known. By contrast, when no such knowledge is given, it 
might be 
better to keep $\gamma$ (or $\beta$) free.

The second ambiguity is that the role of $s=u_1$ is interchangeable with 
any 
independent component of $\x$, that is, with any $u_i$, $i=2,\dots,d$. This 
fact is known as the permutation problem \cite{cardoso1998,sawada2004sap} 
in 
BSS. As 
was already stated, in this work we assume that an initial guess of either 
${\bf 
	a}$ or ${\bf w}$ is given. 

\subsection{Statistical Model}
The main principle of ICE is the same as that of ICA. We make the 
assumption 
that $s$ and ${\bf z}$ are {\em independent}, and ICE is formulated as 
follows:
\begin{quote}
	{\em Find vectors ${\bf a}$ and ${\bf w}$ such that ${\bf w}^T\x$ and 
	${\bf 
			B}\x$ are 	independent (or as close to independent as 
			possible).}
\end{quote}

Let the pdf of $s$ and of $\z$ be, respectively, denoted by $p_s(\xi_1)$ 
and 
$p_\z(\bxi_2)$; $\xi_1$ and $\bxi_2$ denote free variables of appropriate 
dimensions. The joint pdf of $s$ and $\z$ is, owing to their mutual 
independence,
\begin{equation}\label{statmodelICA}
p_\s(\bxi) = p_s(\xi_1)\cdot p_\z(\bxi_2),
\end{equation}
where $\bxi = [\xi_1;\bxi_2]$.
From \eqref{demixing}, the joint pdf of the mixed signals $\x=\A_{\rm 
ICE}{\bf 
	v}$ is
\begin{align}
p_\x(\bxi) &= p_s({\bf w}^H\bxi)\cdot p_\z({\bf B}\bxi)\cdot |\det{\W_{\rm 
		ICE}}|^2 \\
&= p_s(\overline{\beta}\xi_1 + {\bf h}^H\bxi_2)\cdot p_\z(\xi_1{\bf 
	g}-\gamma\bxi_2)\cdot|\gamma|^{2(d-2)},
\end{align}
where the identity
\begin{align}\label{detWICE}
\det{\W_{\rm ICE}}&=(-1)^{d-1}\gamma^{d-2}\\
&=(-1)^{d-1}\beta^{-(d-2)}(1-\h^H\g)^{d-2},
\end{align}
was used, which can easily be verified from \eqref{demixing} using  
\eqref{betagamma}.

The log-likelihood function of $N$ signal samples 
depends on ${\bf a}$ and ${\bf w}$; hence it is
\begin{align}
\mathcal{L}({\bf a}, {\bf w})
&= \frac{1}{N}\log \prod_{n=1}^N p_\x({\bf a},{\bf w}|\x(n))\\\nonumber
&= \frac{1}{N}\sum_{n=1}^N \log p_s({\bf w}^H\x(n))+
\frac{1}{N}\sum_{n=1}^N\log p_\z({\bf B}\x(n))\\\label{loglik}
&+(d-2)\log |\gamma|^2.
\end{align}

\subsection{Gaussian Background}
As previously explained in Section~\ref{sec:mixing}, the background signals 
are 
highly probable to
remain unmixed after ICE, unless $d=2$. This opens the 
problem of modeling the pdf of $\z$. 
A straightforward choice is that the components of $\z$ have the circularly 
symmetric Gaussian 
distribution with zero mean and covariance $\Cz$, i.e., 
$\z\sim\mathcal{CN}({\bf 0},\Cz)$. This choice can be justified by the fact 
that the said components are mixed and correlated; moreover, from the 
Central 
Limit Theorem 
it follows that their distribution is close to Gaussian 
\cite{cardoso1998}. The covariance matrix $\Cz={\rm E}[\z\z^H]$ is
a nuisance parameter.

In this paper, we restrain our considerations to this Gaussian background 
model, noting that other choices are worthy of future investigation. Hence, 
\eqref{loglik} takes the form
\begin{multline}\label{gauloglik}
\mathcal{L}({\bf a}, {\bf w})
= \frac{1}{N}\sum_{n=1}^N \log p_s({\bf w}^H\x(n))-
\frac{1}{N}\sum_{n=1}^N\x(n)^H{\bf B}^H\Cz^{-1}{\bf B}\x(n)
+(d-2)\log |\gamma|^2 \\  -\log\det{\Cz} - d\log\pi.
\end{multline}

\subsection{Orthogonal Constraint}
By inspecting \eqref{loglik} and \eqref{gauloglik}, it can be seen that the 
link between ${\bf a}$ and ${\bf w}$, which are both related 
to the SOI, is rather ``weak''. Indeed, the first term on the 
right-hand side of \eqref{loglik} depends purely on ${\bf w}$, while the 
second 
and the third terms depend purely on ${\bf a}$. The only link between 
${\bf a}$ and ${\bf w}$ is thus expressed in \eqref{betagamma}. 
Consequently, 
the 
log-likelihood function can have spurious maxima where ${\bf a}$ and ${\bf 
w}$ 
do not jointly correspond to the SOI. 

Many ICA algorithms impose the orthogonal constraint (OG) 
\cite{cardoso1994}, 
which decreases the number of unknown parameters in the mixing model. 
This constraint can be used to avoid spurious solutions in ICE and to 
stabilize the convergence of algorithms. Let now $\W_{\rm ICE}$ denote an 
ICE 
de-mixing matrix estimate 
and
\begin{equation}\label{demixedsignals}
\widehat{\bf V}=\begin{pmatrix}
\widehat{{\bf s}} \\ \widehat{\bf Z}
\end{pmatrix}=\W_{\rm ICE}\X
\end{equation}
be the estimated de-mixed signals{\color{black}, that is, $\widehat{{\bf 
s}}$ be the $1\times N$ row vector of samples of the extracted SOI, and 
$\widehat{\bf Z}$ be the $(d-1)\times N$ matrix of samples of the 
background signals.} The OG means that
\begin{equation}\label{orthconst}
\frac{1}{N}\widehat{{\bf s}}\cdot \widehat{\bf Z}^H=\frac{1}{N}{\bf 
	w}^H\X\X^H{\bf B}^H={\bf 
	w}^H\widehat\Cx{\bf 
	B}^H={\bf 0},
\end{equation}
where $\widehat\Cx=\X\X^H/N$ is the sample-based estimate of $\Cx={\rm 
	E}[\x\x^H]$. 

The OG introduces a link between $\a$ and ${\bf w}$, so $\W_{\rm ICE}$ is 
a function of either $\a$ or ${\bf w}$. The dependencies, whose derivations 
are 
given in 
Appendix~A, are
\begin{equation}\label{couplingw}
{\bf w}=\frac{\widehat\Cx^{-1}{\bf 
		a}}{{\bf a}^H\widehat\Cx^{-1}{\bf a}},
\end{equation}
when ${\bf a}$ is the dependent variable, and
\begin{equation}\label{couplinga}
{\bf a}=\frac{\widehat\Cx{\bf w}}{{\bf w}^H\widehat\Cx{\bf w}},
\end{equation}
when the dependent variable is ${\bf w}$.

Interestingly, the coupling \eqref{couplingw} corresponds to the 
approximation 
of
\begin{equation}\label{MPDR}
{\bf w}_{\rm MPDR}^H\x = \frac{{\bf a}^H\Cx^{-1}}{{\bf a}^H\Cx^{-1}{\bf 
a}}\x,
\end{equation}
which is the minimum-power distortionless (MPDR) beamformer steered in the 
direction given by ${\bf a}$, the well-known optimum beamformer in array 
processing theory \cite{vantrees2002}; see also \cite{koldovsky2017b}. In 
Appendix~B, it is shown that if ${\bf a}$ is equal to its true value, then 
\begin{equation}\label{optimumMPDR}
{\bf w}_{\rm MPDR}^H\x = s.
\end{equation}

The advantage of \eqref{couplinga} is that the computation of ${\bf a}$ 
does 
not involve the inverse of $\widehat\Cx$.

\section{Gradient-Based ICE Algorithms}\label{sec:algorithms}
In this section, we derive gradient ICE algorithms aiming at the maximum 
likelihood estimation through searching for the maximum of 
\eqref{gauloglik}. 
Since $p_s$ and $\Cz$ in \eqref{gauloglik} are not known, we propose a 
contrast 
function replacing the true one where $p_s$ and $\Cz$ are 
approximated in a certain way. This is sometimes referred to as the 
quasi-maximum 
likelihood approach; see, e.g., \cite{pham1997}.

%{\color{black} We present two variants constrained by the OG performing 
%the 
%optimization, 
%respectively, in variable ${\bf w}$ and ${\bf a}$. The former variant 
%shows 
%stable and fast convergence when the input ISR is ``low'', while the 
%latter 
%variant has exactly the opposite qualities. Based on this, we propose a 
%novel 
%method that automatically selects the better optimization strategy.}

\subsection{Optimization in ${\bf w}$}

For the optimization in ${\bf w}$, given the coupling \eqref{couplinga}, 
$\beta$ is selected as a free variable while $\gamma$ is dependent. 
Following 
\eqref{gauloglik}, the contrast function is defined as
\begin{align}
\mathcal{C}(\a,{\bf w}) 
&= \frac{1}{N}\sum_{n=1}^N \Bigl\{\log f({\bf w}^H\x(n))
-\x(n)^H{\bf 
	B}^H\R{\bf B}\x(n)\Bigr\} +(d-2)\log |\gamma|^2,\label{contrastcon}
\end{align}
where $f(\cdot)$ is the model pdf of the target signal (replacing $p_s$), 
and 
${\bf R}$ is a weighting positive definite matrix (replacing $\Cz^{-1}$).

Using the Wirtinger calculus \cite{brandwood1983,li2008}, we derive in 
Appendix~C that the gradient of 
$\mathcal{C}$ with respect to ${\bf w}^H$, under the 
coupling \eqref{couplinga}, equals
\begin{multline}\label{grad1}
\left.\frac{\partial\mathcal{C}}{\partial{\bf 
		w}^H}\right|_\text{w.r.t. \eqref{couplinga}} = 
-\frac{1}{N}\X\phi({\bf w}^H\X)^T + 2\a\,\tr(\R\wCz)
-({\bf w}^H\wCx{\bf w})^{-1}\bigl(\wCx{\bf E}^H\R\wCz\h-\tr(\R{\bf 
B}\wCx{\bf 
	E}^H)\wCx{\bf e}_1\bigr)\\
-2(d-2)\a+\overline{\gamma}^{-1}(d-2)({\bf w}^H\wCx{\bf w})^{-1}\wCx{\bf 
e}_1,
\end{multline}
where $\tr(\cdot)$ denotes the trace; ${\bf E}=[{\bf 
	0}\quad\I_{d-1}]$; ${\bf e}_1$ denotes the first column of $\I_d$; and
\begin{equation}\label{score}
\phi(\xi) = -\frac{\partial \log f(\xi)}{\partial \xi}
\end{equation}
is the score function of the model pdf $f(\cdot)$. 

Now, we put $\R=\wCz^{-1}$; this is a choice for which the derivative of 
\eqref{gauloglik} with respect to the unknown parameter $\Cz$ is equal to 
zero. 
Then, the following 
identities can be applied in \eqref{grad1}.
\begin{align}
\tr(\wCz^{-1}\wCz)&=\tr(\I_{d-1})=d-1,\\
{\bf E}^H\h+\beta{\bf e}_1&={\bf w},\\
\wCz^{-1}{\bf B}\wCx&=\wCz^{-1}{\bf B}\X\X^H/N\nonumber\\
&=\wCz^{-1}\widehat{\bf 
	Z}[\widehat{\bf s}^H\,\,\,\widehat{\bf Z}^H]\A_{\rm ICE}^H={\bf 
	E}\A_{\rm 
	ICE}^H\\
\tr(\wCz^{-1}{\bf B}\wCx{\bf E}^H)&=\tr({\bf E}\A_{\rm ICE}^H{\bf 
	E}^H)=\nonumber\\&=\overline{\gamma}^{-1}\tr(\h\g^H-{\bf 
	I}_{d-1})=\nonumber\\
&=-\beta-(d-2)\overline{\gamma}^{-1},
\end{align}
where we used \eqref{demixedsignals} and \eqref{orthconst};
\eqref{grad1} is now simplified to
\begin{equation}\label{grad}
\left.\frac{\partial\mathcal{C}}{\partial{\bf 
		w}^H}\right|_\text{w.r.t. 
	\eqref{couplinga}} = 
\a-\frac{1}{N}\X\phi({\bf w}^H\X)^T.
\end{equation}

In fact, $\R=\wCz^{-1}$ depends on the current value of ${\bf w}$ since 
$\wCz={\bf B}\wCx{\bf B}^H$. It means that, with any estimate 
of ${\bf w}$, the distribution of $\widehat{\bf Z}={\bf B}\X$ is assumed to 
be 
$\mathcal{CN}({\bf 0},\wCz)$, which obviously introduces little (or no) 
information into the contrast function. $\wCz$ is close to the true 
covariance 
$\Cz$ only when ${\bf a}$ is close to its true value.

For $N\rightarrow+\infty$, \eqref{grad} takes on the form
\begin{equation}\label{gradE}
\frac{\partial\mathcal{C}}{\partial{\bf w}^H} = 
\a-{\rm E}[\x\phi({\bf w}^H\x)].
\end{equation}
When ${\bf w}$ is the ideal separating vector, that is, when ${\bf 
w}^H\x=s$, 
then from \eqref{modelsinglesource} it follows that 
\begin{equation}\label{gradE2}
\frac{\partial\mathcal{C}}{\partial{\bf w}^H} = 
(1-{\rm E}[s\phi(s)])\a.
\end{equation}
This shows us that the true separating vector is a stationary point of the 
contrast function only if $\phi(\cdot)$ satisfies the condition
\begin{equation}\label{condition}
{\rm E}[s\phi(s)]=1.
\end{equation}

Based on this observation, we propose a method whose steps are described 
in Algorithm~\ref{algorithm1}. In every step, it iterates in the direction 
of 
the steepest ascent of $\mathcal{C}$ where $\R=\wCz^{-1}$ (step 9), and 
$\phi(\cdot)$ is 
normalized 
so that condition \eqref{condition} is satisfied for the current target 
signal 
estimate $\widehat{s}={\bf w}^H{\bf x}$, {\color{black} that is,  
$\phi(\widehat{s})\leftarrow\phi(\widehat{s})/(\widehat{s}\phi(\widehat{s})^T/N)$
(see steps~7 and 8).} 
This is repeated until 
the norm of the gradient is 
smaller than ${\tt tol}$; $\mu$ is the step length parameter; and ${\bf 
w}_{\rm 
	ini}$ is the initial guess. We call this method OGICE$_{\bf w}$.

%{\color{black} When the observations $\X$ are preprocessed (whitened) so 
%that $\wCx=\I_d$, steps 4 and 5 in Algorithm~\ref{algorithm1} are 
%simplified to
%\[
%\a\leftarrow {\bf w}/\|{\bf w}\|^2.
%\]}

\begin{algorithm}\label{algorithm1}
	\caption{OGICE$_{\bf w}$: separating vector estimation based on 
		orthogonally constrained gradient-ascent algorithm}
	\SetAlgoLined
	\KwIn{$\X$, ${\bf w}_{\rm ini}$, $\mu$, ${\tt tol}$}
	\KwOut{${\bf a},{\bf w}$}
	$\widehat\C_\x=\X\X^H/N$;\\
	${\bf w}={\bf w}_{\rm ini}$;\\
	\Repeat{$\|\Delta\|<{\tt tol}$}{
		$\lambda_{\bf w} \leftarrow ({\bf w}^H\widehat\C_\x{\bf w})^{-1}$;\\
		${\bf a} \leftarrow \lambda_{\bf w}\widehat\C_\x{\bf 
		w}$;\tcc*[f]{OG 
			constraint \eqref{couplinga}}\\
		$\widehat{\bf s} \leftarrow {\bf w}^H\X$;\\
		$\nu \leftarrow \widehat{\bf s}\phi(\widehat{\bf 
		s})^T/N$;\tcc*[f]{due 
			to cond. \eqref{condition}}\\
		$\Delta \leftarrow {\bf a}-\nu^{-1}\X\phi(\widehat{\bf 
			s})^T/N$;\tcc*[f]{by \eqref{grad}}\\
		${\bf w} \leftarrow {\bf w} + \mu\Delta$;\tcc*[f]{gradient ascent}\\
	}
\end{algorithm}

In fact, \eqref{gradE} coincides with the gradient of a heuristic criterion 
derived from mutual information in \cite{pham2006} (page 870, Eq. 4). The 
author, D.-T. Pham, called this approach ``Blind Partial Separation''. Our 
derivation provides a deeper insight into this result by showing its 
connection 
with maximum likelihood estimation. Most 
importantly, it is seen that \eqref{grad} follows from a particular 
parameterization of the (de)-mixing model, it imposes the OG between ${\bf 
a}$ 
and ${\bf w}$, and it relies on the Gaussian modeling of the background 
signals 
whose covariance is estimated as $\wCz$.

\subsection{Optimization in ${\bf a}$}
The gradient with respect to ${\bf a}$ when ${\bf w}$ is dependent through 
\eqref{couplingw} and when $\gamma=1$ has been derived in 
\cite{koldovsky2017b}. Treating $\gamma$ as a free 
variable, and by putting $\R=\wCz^{-1}$, the gradient reads
\begin{equation}\label{gradw}
\left.\frac{\partial\mathcal{C}}{\partial{\bf 
		a}^H}\right|_\text{w.r.t. 
	\eqref{couplingw}} = 
{\bf w}-\frac{\lambda_{\bf a}}{N}\wCx^{-1}\X\phi({\bf w}^H\X)^T,
\end{equation}
where $\lambda_{\bf a} = (\a^H\wCx^{-1}\a)^{-1}$. For 
$N\rightarrow+\infty$, 
the true mixing vector is a stationary point only if \eqref{condition} is 
fulfilled. The corresponding algorithm, similar to that proposed in 
\cite{koldovsky2017b} but leaving $\gamma$ free, will be referred to as 
OGICE$_{\bf a}$. %, is described in Algorithm~\ref{algorithm2}.

{\color{black}
	\subsection{Preconditioning}
	\label{subsec:Preconditioning}
	The multiplicative form of the mixing model \eqref{mixingICE} 
	allows us to consider the gradient computed according to transformed 
	input 
	signals ${\bf U}={\bf D}\X$, where ${\bf D}$ is a {\em preconditioning} 
	non-singular 
	matrix. We will consider the preconditioning applied within 
	OGICE$_{\bf w}$ since this will help us to reveal the connection 
	between 
	OGICE$_{\bf w}$ and three well-known ICA/BSE algorithms.
	
	Let ${\bf w}_{\bf x}$ and ${\bf w}_{\bf u}$ be the separating vectors 
	operating 
	on $\X$ and ${\bf U}$, respectively, giving the same extracted signal, 
	i.e., 
	$\widehat{\bf s}={\bf w}_{\bf x}^H\X={\bf w}_{\bf u}^H{\bf U}$. It 
	follows that 
	${\bf w}_{\bf x}={\bf D}^H{\bf w}_{\bf u}$.  
	Consider now the gradient \eqref{grad} when the input data are ${\bf 
	U}$ 
	and the initial vector is ${\bf w}_{\bf u}$, which will be denoted by 
	$\Delta_{\bf u}$. The sample covariance matrix of 
	${\bf U}$ is $\widehat{\bf C}_{\bf u}={\bf D}\wCx{\bf D}^H$, so the
	right-hand side of \eqref{grad} gives
	\begin{multline}\label{modgrad}
	\Delta_{\bf u}=\frac{\widehat{\bf C}_{\bf u}{\bf w}_{\bf u}}{{\bf 
	w}_{\bf 
			u}^H\widehat{\bf C}_{\bf u}{\bf w}_{\bf u}}-\frac{1}{N}{\bf 
		U}\phi(\widehat{\bf s})^T=
	\frac{{\bf D}\wCx{\bf D}^H{\bf w}_{\bf u}}{{\bf w}_{\bf u}^H{\bf 
	D}\wCx{\bf 
			D}^H{\bf w}_{\bf u}}-\frac{1}{N}{\bf D}\X\phi(\widehat{\bf 
		s})^T=\\
	{\bf D}\left(\frac{\wCx{\bf w}_{\bf x}}{{\bf w}_{\bf x}^H\wCx{\bf 
	w}_{\bf 
			x}}-\frac{1}{N}\X\phi(\widehat{\bf s})^T\right)={\bf 
			D}\Delta_{\bf x},
	\end{multline}
	where $\Delta_{\bf x}$ denotes the ``normal'' gradient, that is, when 
	the input 
	data are ${\bf X}$ and the initial vector is ${\bf w}_{\bf x}$. Note 
	that 
	\eqref{modgrad} remains valid when the normalization of $\phi$ 
	(dividing by 
	$\nu$) is taken into account, because $\nu$ is only a function of 
	$\widehat{s}$.
	
	After ${\bf w}_{\bf u}$ is updated as ${\bf w}_{\bf u}^{\rm new}= {\bf 
		w}_{\bf u}+\mu\Delta_{\bf u}$, 
	the extracted signal is equal to
	\begin{multline}
	({\bf w}_{\bf u}^{\rm new})^H{\bf U}=\left({\bf w}_{\bf 
	u}+\mu\Delta_{\bf 
		u}\right)^H{\bf U}=
	\left({\bf D}^{-H}{\bf w}_{\bf x}+\mu{\bf D}\Delta_{\bf x}\right)^H{\bf 
	DX}=
	\left({\bf w}_{\bf x}+\mu{\bf D}^H{\bf D}\Delta_{\bf x}\right)^H{\bf X}.
	\end{multline}
	It follows that the gradient update computed on the preconditioned data 
	${\bf 
		U}$ corresponds with a modified update rule for ${\bf w}_{\bf x}$ 
		given by
	\begin{equation}\label{relgrad}
	{\bf w}_{\bf x}\leftarrow{\bf w}_{\bf x}+\mu{\bf D}^H{\bf D}\Delta_{\bf 
	x}.
	\end{equation}
	
	For ${\bf D}={\bf I}_d$, the modified update rule obviously coincides 
	with the 
	original one. In the following subsection, we will consider other 
	special 
	choices of ${\bf 
		D}$ and compare the modified OGICE$_{\bf w}$ with other ICA/BSE 
		methods known 
	in the literature.

	\subsection{Relation to gradient and natural gradient ICA methods}
	Here, OGICE$_{\bf w}$ is compared with the method by Bell and 
	Sejnowski \cite{bell1995} for ICA and with its popular modification 
	known as 
	Amari's Natural Gradient (NG) algorithm \cite{amari1996}; see also 
	\cite{laheld1996} and \cite{li2008} for the complex-valued variant. In 
	each 
	step of the Bell and Sejnowski's method (BS), the whole de-mixing 
	matrix is 
	updated as
	\begin{equation}
	\Delta\W\leftarrow \W^{-H}- \overline{\phi({\bf W}\X)}\X^H/N.
	\end{equation}
	After taking the conjugate transpose on both sides, and denoting 
	$\W^{-1}=\A$, 
	this update can be re-written as
	\begin{equation}\label{ICAupdate}
	\Delta\W^H\leftarrow \A - \X\phi({\bf W}\X)^T/N.
	\end{equation}
	Now, the right-hand side of \eqref{grad} corresponds to any row on the 
	right-hand side of \eqref{ICAupdate}. 
	
	The de-mixing matrix update in NG is obtained when the right-hand side 
	of 
	\eqref{ICAupdate} is multiplied by $\W^H\W$ from left, which gives
	\begin{equation}\label{NGupdate}
	\Delta\W^H\leftarrow {\bf W}^H\bigl({\bf I}_d - \W\X\phi({\bf 
		W}\X)^T/N\bigr).
	\end{equation}
	OGICE$_{\bf w}$ becomes similar to NG when considering it with the 
	modified 
	update \eqref{relgrad} where the precondition matrix ${\bf D}={\bf 
	W}_{\rm 
		ICE}$. This choice corresponds to the update when the input data 
		are 
	pre-separated by the current de-mixing matrix prior to each 
	iteration, and the starting ${\bf w}$ is equal to the unit vector (the 
	first 
	column of ${\bf I}_d$).

	The main difference between OGICE$_{\bf w}$ and the respective ICA 
	algorithms 
	thus resides in that BS and NG perform updates of the whole de-mixing 
	matrix, while OGICE$_{\bf w}$ updates only its first row (the 
	separating 
	vector) under the orthogonal constraint. Next, the nonlinearity in 
	OGICE$_{\bf 
		w}$ is normalized according to \eqref{condition}, while neither BS 
		nor NG apply 
	any normalization.

	%From this perspective, OGICE$_{\bf w}$ appears to be most 
	%similar to the Scaled Natural Gradient algorithm (NGsc) proposed in 
	%\cite{douglas2007}, which is popular due to faster convergence as 
	%compared to 
	%the original NG.
	
	\subsection{Relation to One-unit FastICA}\label{sec:fastica}
	One-unit FastICA (FICA) was derived as a fixed-point algorithm that 
	minimizes 
	the entropy of the extracted signal under the unit scale constraint. 
	The FICA 
	update for the separating vector can be written as \cite{hyvarinen1997}
	\begin{align}\label{fastica}
	{\bf w} &\leftarrow {\bf w}-\bigl(\wCx^{-1}{\bf X}\psi(\widehat{\bf 
		s})^T/N-\nu{\bf w}\bigr)/(\rho-\nu),\\
	{\bf w} &\leftarrow {\bf w}/\sqrt{{\bf w}^H\wCx{\bf 
	w}}\label{unitscale},
	\end{align}
	where $\nu=\widehat{\bf s}\psi(\widehat{\bf s})^T/N$ and 
	$\rho=\psi'(\widehat{\bf s}){\bf 1}_N/N$, where $\psi'$ is the 
	derivative of 
	$\psi$, and ${\bf 1}_N$ denotes the column vector of ones of length 
	$N$. Note 
	that \eqref{unitscale} corresponds to the normalization of ${\bf w}$ so 
	that 
	the scale of the extracted signal is one. 
	
	FICA is more known when it operates on pre-whitened data $\X$, which 
	means that 
	they are normalized prior to the optimization so that their sample 
	covariance 
	matrix is ${\bf I}_d$. This corresponds with the choice of the 
	preconditioning 
	matrix in 
	Section~\ref{subsec:Preconditioning} as ${\bf D}={\bf F}\wCx^{-1/2}$, 
	where 
	$\wCx^{-1/2}$ denotes the inverse matrix square root of $\wCx$, and 
	${\bf F}$ 
	is an 
	arbitrary unitary matrix. Then, it holds that ${\bf D}^H{\bf 
	D}=\wCx^{-1}$, and 
	we can compare the modified update rule of OGICE$_{\bf w}$ with 
	\eqref{fastica}. Specifically, the OGICE$_{\bf w}$ update modified 
	according to 
	\eqref{relgrad} together with the nonlinearity normalization can be 
	written as
	\begin{equation}\label{relogice}
	{\bf w}\leftarrow {\bf w}+\mu\left(\frac{{\bf w}}{{\bf w}^H\wCx{\bf 
			w}}-\nu^{-1}\frac{1}{N}\wCx^{-1}\X\phi(\widehat{\bf s})^T\right)
	\end{equation}
	where $\mu$ is the step length parameter. By comparing \eqref{fastica} 
	and 
	\eqref{relogice}, the updates coincide when $\mu=\frac{\nu}{\rho-\nu}$ 
	provided that ${\bf w}^H\wCx{\bf w}=1$. 
	
	In conclusion, FICA and OGICE$_{\bf w}$ correspond to the same method 
	when (a) 
	the input data are pre-whitened (directly or through the 
	preconditioning 
	matrix and the modified update), (b) the step length in OGICE$_{\bf w}$ 
	is 
	selected adaptively as 
	$\mu=\frac{\nu}{\rho-\nu}$, and (c) OGICE$_{\bf w}$ is forced to 
	operate on 
	the unit-scale sphere, which can be achieved through normalizing ${\bf 
	w}$ 
	after each iteration as in \eqref{unitscale}. These results extend the 
	analysis done in \cite{hyvarinen1999b}.

	%\begin{algorithm}[t]
	%	\label{algorithm2}
	%	\caption{OGICE$_{\bf a}$: mixing vector estimation based on 
	%orthogonally 
	%	constrained gradient-ascent algorithm}
	%	\SetAlgoLined
	%	\KwIn{$\X$, ${\bf a}_{\rm ini}$, $\mu$, ${\tt tol}$}
	%	\KwOut{${\bf a},{\bf w}$}
	%	$\widehat\C_\x=\X\X^H/N$;\\
	%	${\bf a}={\bf a}_{\rm ini}$;\\
	%	\Repeat{$\|\Delta\|<{\tt tol}$}{
	%		$\lambda_{\bf a} \leftarrow ({\bf a}^H\widehat\C_\x^{-1}{\bf 
	%			a})^{-1}$;\\
	%		${\bf w} \leftarrow \lambda_{\bf a}\widehat\C_\x^{-1}{\bf 
	%			a}$;\tcc*[f]{OG constraint \eqref{couplingw}}\\
	%		$\widehat{\bf s} \leftarrow {\bf w}^H\X$;\\
	%		${\bf q} \leftarrow \phi(\widehat{\bf s})/(\widehat{\bf 
	%			s}\phi(\widehat{\bf s})^T/N)$;\tcc*[f]{to fulfil 
	%			\eqref{condition}}\\
	%		$\Delta \leftarrow {\bf w}-\lambda_{\bf 
	%			a}\widehat\C_\x^{-1}\X{\bf q}^T/N$;\tcc*[f]{by 
	%			\eqref{gradw}}\\
	%		${\bf a} \leftarrow {\bf a} + \mu\Delta$;\tcc*[f]{gradient 
	%ascent}\\
	%	}
	%\end{algorithm}
	
	\subsection{Switched optimization}\label{sec:switchedopt}
	
	When all sources should be separated, as in ICA, it is less important 
	which 
	source is extracted in which output channel as all sources are finally 
	separated. However, when only one source should be extracted (based on 
	the 
	initial value of the mixing/separating vector), the size of the region 
	of 
	convergence (ROC) to the SOI becomes essential.
	
	The ROC depends on the surface of the objective function of the given 
	algorithm. This is influenced by all properties of the observed 
	signals, 
	namely, by the signals' distributions and by the initial 
	Signal-to-Interference 
	Ratio (SIR), where the latter is a function of the signals' scales and 
	of the 
	mixing matrix. The influence of the initial SIR is, however, difficult 
	to 
	analyze as it is different on each input channel. 
	
	Nevertheless, we can constrain our considerations to situations where 
	the 
	initial SIR is approximately the same on all channels. This happens, 
	for 
	example, when mutual distances of sensors are small compared to the 
	distances 
	of the sources from the sensor array. Then, we can assume that the 
	magnitude of 
	each element of the mixing matrix is approximately equal to a constant, 
	so the 
	initial SIRs are mainly influenced by the scales of the sources. % SOI 
	%as it 
	%compares to the average scale of the other sources. 
	Let us define Scales Ratio (SR) related 
	to the SOI as
	\begin{equation}\label{SIRini}
	{\rm SR}=\frac{{\rm E}[|s|^2]}{\frac{1}{d-1}\sum_{i=2}^d{\rm 
	E}[|u_i|^2]}.
	\end{equation}
	%although a more precise definition also depends on the mixing matrix 
	%and the 
	%selected input channel. 
	The following example shows how the ROC of the OGICE algorithms can be 
	influenced by ${\rm SR}$.

	Consider a situation where the SOI is a ``weak'' signal, i.e., ${\rm 
	SR}\ll 
	0$~dB. The mixing vector ${\bf a}$ is then ``hard'' to find while the 
	background subspace can be identified ``easily''. For the de-mixing 
	matrix, the problem is reciprocal. The estimation of ${\bf B}$ in 
	\eqref{demixing} is inaccurate as ${\bf B}$ depends purely on ${\bf 
	a}$, while 
	the estimation of ${\bf w}$ yields a low variance. %${\rm SR}$ 
	%therefore has
	%significant influence on the shape of the contrast function as well as 
	%on the 
	%size of the region of convergence (ROC) in the vicinity of the 
	%solution. 

	Fig.~\ref{fig:contrast} shows an objective function in the case of a 
	real-valued mixture of two Laplacean components where one plays the 
	role of the 
	SOI and the other one is the background source (but the roles can be 
	interchanged); the number of samples is $N=1000$; the mixing matrix is 
	$\A=\left(\begin{smallmatrix}
	1&1\\-1&1
	\end{smallmatrix}\right)$; ${\rm SR}$ is considered in two settings: 
	0~dB and 
	-20~dB, respectively. The function \eqref{contrastcon} is shown as it 
	depends 
	on $a$ and $w$, respectively, where the mixing vector is ${\bf 
	a}=[1;a]$ and 
	the separating vector is ${\bf w}=[1;w]$, respectively. The 
	nonlinearity is 
	$\log f(x)=-\log|\cosh(x/\sigma_x)|$ where $\sigma_x^2$ is the 
	variance of the input. The perfect extraction of the SOI is achieved 
	for $a=-1$ 
	and $w=-1$, while $a=1$ and $w=1$ correspond to the extraction of the 
	background source.

	\begin{figure}[t]
		\centering
		\includegraphics[width=0.9\linewidth]{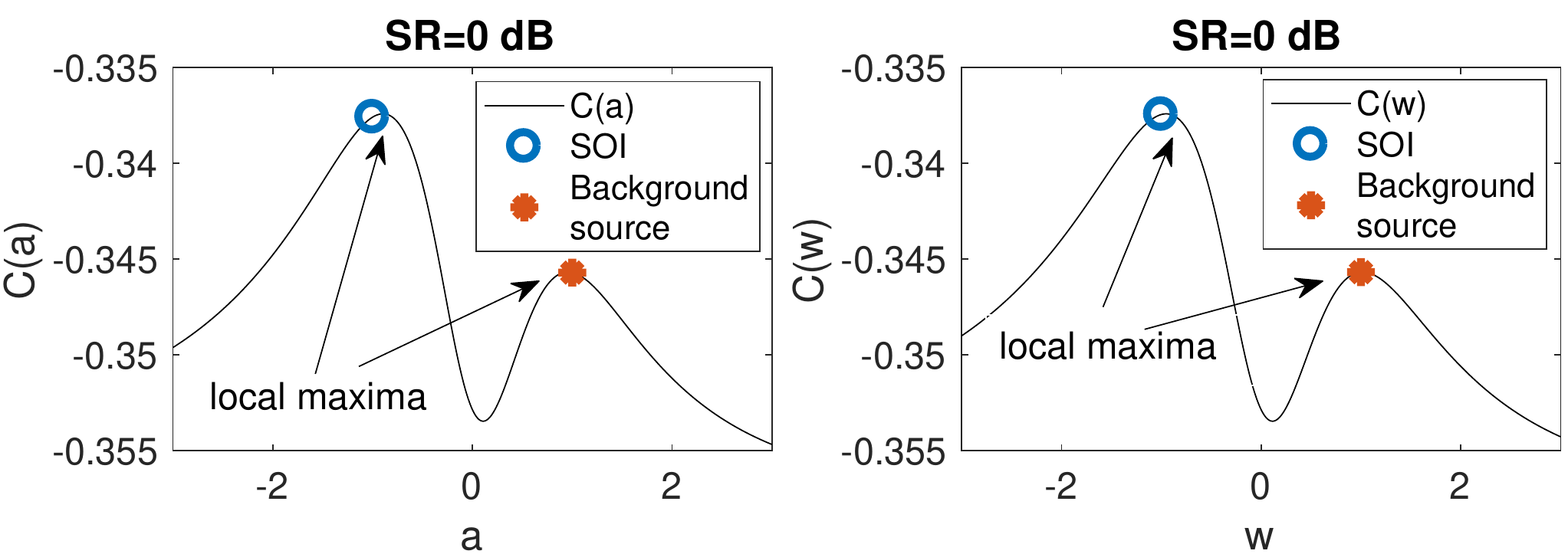}\\
		\includegraphics[width=0.9\linewidth]{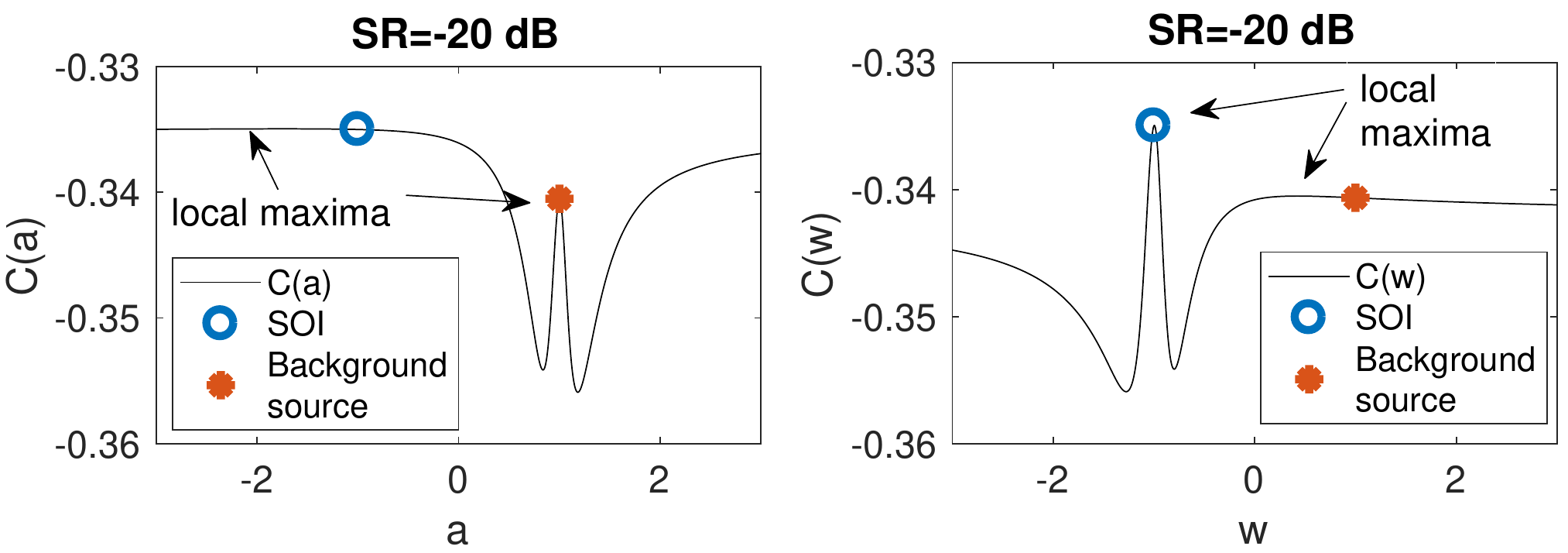}
		\caption{Examples of the contrast function \eqref{contrastcon} as 
		it 
			depends on ${\bf a}=[1;a]$ and ${\bf w}=[1;w]$, respectively, 
			when 
			${\rm SR}$ is 0~dB and -20~dB.   
			\label{fig:contrast}}
	\end{figure}
	
	For ${\rm SR}=0$~dB, the surface of the contrast function as it depends 
	on $a$ 
	or $w$, respectively, is almost the same. The local maxima are slightly 
	biased 
	from perfect solutions, and the sizes of ROC related to the maxima are 
	approximately equal for both sources. 
	
	For ${\rm SR}=-20$~dB, the background source dominates the mixture. 
	Here, the 
	maximum corresponding to the mixing vector of the SOI is significantly 
	biased 
	from its ideal value, the function is almost flat in the vicinity of 
	that 
	maximum, and the corresponding ROC is wide. By contrast, the separating 
	vector for SOI is precisely localized by a sharp local maximum, which 
	has 
	a narrow ROC. The exact opposite is true for the maxima corresponding 
	to the 
	dominating background source.

	In this example, OGICE$_{\bf w}$ is more advantageous when SR$_{\rm 
	in}\gg 
	0$~dB in the sense that the ROC corresponding to the SOI is wide. Even 
	when the 
	initialization of OGICE$_{\bf w}$ is significantly deviated, the 
	probability of 
	the successful convergence is high. 
	%When OGICE$_{\bf w}$ is 
	%initialized within the ROC, its convergence is faster as the gradient 
	%is steep 
	%enough. 
	Similarly, OGICE$_{\bf a}$ is advantageous when SIR$_{\rm in}\ll 0$~dB. 
	However, it is worth emphasizing that these properties of the 
	algorithms depend 
	on the mixing matrix, hence, also on the choice of the preconditioning 
	matrix 
	introduced in Section~\ref{subsec:Preconditioning}. For example, 
	OGICE$_{\bf 
		w}$ with the modified update \eqref{relgrad} is advantageous in the 
		example 
	here when ${\bf D}={\bf I}_d$.
	
	%These phenomena may not be visible through ICA methods, which estimate 
	%all 
	%sources (dominant and week) simultaneously.
	
	The above observations suggest that partial knowledge of mixing 
	conditions 
	can be used towards successful extraction of the SOI.}
In a purely blind scenario, ${\rm SR}$ is not known. We therefore propose 
the 
following heuristic approach for the selection between the optimization in 
${\bf a}$ and ${\bf w}$. Let $\a$ be the current estimate of the mixing 
vector. 
Then,
\begin{equation}
{\bf b} = \wCx\a
\end{equation}
can be viewed as an approximate eigenvector of $\wCx$ corresponding to 
the largest eigenvalue, which is approximately equal to $\lambda_{\bf 
	b}=b_1/a_1$. Thus, $\bigl\|{\bf a}/a_1-{\bf b}/\lambda_{\bf b}\bigr\|$ 
	is small 
when $\a$ is close to the dominant eigenvector.

Next, to assess the dominance of that eigenvector, that is, whether it is 
significantly larger compared to the other eigenvectors, we propose to 
compute 
the ratio of norms of matrices $\wCx-\lambda_{\bf b}{\bf b}{\bf b}^H/\|{\bf 
	b}\|^2$ and $\wCx$. The ratio is small when ${\bf b}$ is a dominant 
	eigenvector 
of $\wCx$. The criterion of ``proximity and dominance'' is therefore 
defined as
\begin{equation}\label{critoptim}
\mathcal{B}({\bf a})=\bigl\|{\bf a}/a_1-{\bf 
	b}/\lambda_{\bf b}\bigr\| \,\frac{\left\|\wCx-\lambda_{\bf b}{\bf 
		b}{\bf b}^H/\|{\bf b}\|^2\right\|_F}{\|\wCx\|_F}.
\end{equation}

The proposed algorithm, referred to as OGICE$_{\rm s}$, selects the 
optimization parameter based on the current value of 
$\mathcal{B}({\bf a})$. When $\mathcal{B}({\bf a})<\tau$, the optimization 
in 
${\bf w}$ is selected; otherwise, the optimization proceeds in ${\bf a}$. 
Normally, we select $\tau=0.1$. To lower the computational load, the 
criterion 
is recomputed only once after $Q$ iterations ($Q=10$). The stopping 
condition of OGICE$_{\rm s}$ is the same as that in OGICE$_{\bf w}$ or 
OGICE$_{\bf a}$; see the summary in Algorithm~\ref{algorithm3}.

\begin{algorithm}[t]
	\label{algorithm3}
	\caption{OGICE$_{\bf s}$: ICE algorithm with automatic selection of 
		optimization parameter}
	\SetAlgoLined
	\KwIn{$\X$, ${\bf a}_{\rm ini}$, $\mu$, ${\tt tol}$, $Q$}
	\KwOut{${\bf a},{\bf w}$}
	$\widehat\C_\x=\X\X^H/N$; ${\bf a}={\bf a}_{\rm ini}$; $i=0$;\\
	\Repeat{$\|\Delta\|<{\tt tol}$}{
		\If{$i \equiv 0\, ({\rm mod}\, Q)$}{$\kappa=\mathcal{B}({\bf 
				a})$;\tcc*[f]{using 
				\eqref{critoptim}}}
		$i \leftarrow i+1$;\\
		\uIf{$\kappa<0.1$}{Update ${\bf w}$ as in OGICE$_{\bf w}$\\
			${\bf a} \leftarrow ({\bf w}^H\widehat\C_\x{\bf 
				w})^{-1}\widehat\C_\x{\bf w}$;}
		\Else{Update ${\bf a}$ as in OGICE$_{\bf a}$\\
			${\bf w} \leftarrow ({\bf a}^H\widehat\C_\x^{-1}{\bf 
				a})^{-1}\widehat\C_\x^{-1}{\bf 
				a}$;}
	}
\end{algorithm}

\section{Independent Vector Extraction}
\subsection{Definition}
In this section, {\color{black} we extend the ideas in previous sections, 
which were derived for a single mixture,} for joint extraction of an 
independent 
vector component from a set of $K$ instantaneous mixtures {\color{black} 
(each of the same dimension $d$)}
\begin{equation}\label{mixingIVE}
\x^{k}=\A_{\rm ICE}^k{\bf v}^k, \qquad k=1,\dots,K.
\end{equation}
Here, $\A_{\rm ICE}^k$ obeys the same structure as \eqref{mixingICE}, and 
${\bf 
	v}^k=[s^k;{\bf z}^k]$. A joint mixture model can be written as
\begin{equation}\label{mixingIVEjoint}
\mathbbm{x}=\mathbbm{A}_{\rm IVE}\mathbbm{v}. 
\end{equation}
The double-striked font will be used to denote concatenated variables or 
parameters from $K$ mixtures, e.g., $\mathbbm{x}=[{\bf x}^1;\dots;{\bf 
x}^K]$. 
The joint mixing 
matrix obeys $\mathbbm{A}_{\rm IVE}={\tt bdiag}(\A_{\rm 
	ICE}^1,\dots,\A_{\rm ICE}^K)$, where ${\tt bdiag}(\cdot)$ denotes a 
block-diagonal 
matrix with the arguments on the main block-diagonal.

Although the mixtures ${\bf x}^1,\dots,{\bf x}^K$ are algebraically 
independent, similarly to IVA, we introduce a joint statistical model 
{\color{black} (similarly to \eqref{statmodelICA})} given by
\begin{equation}\label{IVEindependence}
p_{\mathbbm{v}}(\mathbbm{v})=p_{\mathbbm{s}}(\mathbbm{s})p_{\mathbbm{z}}(\mathbbm{z}),
\end{equation}
which means that $\mathbbm{s}$ and $\mathbbm{z}$ are independent but the 
elements inside of them can be dependent. Like in the previous section, we 
constrain 
our considerations to the Gaussian background modeling; so it is assumed 
that
\begin{equation}
\mathbbm{z}\sim\mathcal{CN}({\bf 0},\Czz),
\end{equation}
where 
$\Czz={\rm E}[\mathbbm{z}\mathbbm{z}^H]$ whose $ij$th block of dimension 
\mbox{$(d-1)\times (d-1)$} is
$\Cz^{ij}={\rm E}[{\bf z}^i{{\bf z}^j}^H]$; similarly 
$\C_\mathbbm{x}$ as well as the sample-based counterparts 
$\widehat{\C}_\mathbbm{z}$ and $\widehat{\C}_\mathbbm{x}$ are defined.

Similar to \eqref{contrastcon}, the quasi-likelihood contrast function 
following from \eqref{IVEindependence} (for one signal sample), is
\begin{multline}
\mathcal{J}(\mathbbm{w},\mathbbm{a})=\log 
f\bigl(\widehat{s}^1,\dots,\widehat{s}^K\bigr)  
-\sum_{i=1}^K\sum_{j=1}^K{\x^i}^H{{\bf B}^j}^H\R^{ij}{\bf 
	B}^j\x^j+\sum_{k=1}^K\log|\det\W_{\rm ICE}^k|^2,
\end{multline}
where $\widehat{s}^k=({\bf w}^k)^H\x^k$, and $\R^{ij}$ is a weighting 
matrix substituting the $ij$th block of the unknown $\Czz^{-1}$. %; 
%$\mathbbm{w}$ and $\mathbbm{a}$ are defined similarly 
%to $\mathbbm{x}$.

\subsection{Gradient of the contrast function}
After a lengthy computation, which follows steps similar to those described 
in 
Appendix~C 
(we skip the details to save space), the derivative of $\mathcal{J}$ with 
respect to $({{\bf w}^k})^H$ under the constraints {\color{black} (similar 
to \eqref{couplinga})}
\begin{equation}
{\bf a}^k=\frac{\wCx^{kk}{\bf w}^k}{{{\bf w}^k}^H\wCx^{kk}{\bf w}^k}, \quad 
k=1,\dots,K,
\end{equation}
and when $\R^{k\ell}$ is selected as the $k\ell$th block of 
$\widehat{\C}_\mathbbm{z}^{-1}$, reads
\begin{equation}\label{gradIVE}
\frac{\partial \mathcal{J}}{\partial {{\bf w}^k}^H}={\bf 
	a}^k-\frac{1}{N}\X^k\phi^k\bigl(\widehat{\bf s}^1,\dots,\widehat{\bf 
	s}^K\bigr)^T 
+\frac{1}{{{\bf 
			w}^k}^H\wCx^{kk}{\bf 
		w}^k} {\wCx^{kk}{{\bf B}^k}^H\boldsymbol{\epsilon}^k}.
\end{equation} 
Here, $\widehat{\bf s}^k=({\bf w}^k)^H\X^k$,
\begin{equation}
\phi^k(\xi^1,\dots,\xi^K)=-\frac{\partial \log 
f(\xi^1,\dots,\xi^K)}{\partial 
	\xi^k},
\end{equation}
is the score function related to the model joint density $f(\cdot)$ of 
$\mathbbm{s}$ with respect to the $k$th variable, and
\begin{equation}
\boldsymbol{\epsilon}^k=\sum_{\ell=1}^{K}\R^{k\ell}\boldsymbol{\theta}^{\ell
	k},\quad\text{where}\quad
\boldsymbol{\theta}^{\ell k}=\widehat{\bf Z}^\ell(\widehat{\bf s}^k)^H/N.
\end{equation}

By comparing \eqref{grad} with \eqref{gradIVE}, the latter differs only in 
that 
the nonlinearity $\phi^k(\cdot)$ is dependent on the SOIs separated from 
all 
$K$ mixtures, plus the third term that does not occur in \eqref{grad}.

$\boldsymbol{\theta}^{\ell k}$ is the sample correlation between the 
estimated 
SOI in the $k$th mixture and the separated background in the $\ell$th 
mixture, 
and can also be written as $\boldsymbol{\theta}^{\ell k}={\bf 
B}^\ell\wCx^{\ell 
	k}{\bf w}^k$. For $k=1,\dots,K$, $\boldsymbol{\theta}^{k k}={\bf 0}$ 
	due to 
the OG, but, for $k\neq\ell$, $\boldsymbol{\theta}^{\ell k}$ is non-zero in 
general. Therefore, the third term in \eqref{gradIVE} vanishes only 
when $\R^{k\ell}={\bf 0}$ for $k\neq\ell$, $\ell=1,\dots,K$.

Here, a special case is worth considering in which  $\C_\mathbbm{x}={\tt 
	bdiag}(\Cx^{11},\dots,\Cx^{KK})$; in other words, the mixtures 
\eqref{mixingIVE} are uncorrelated {\color{black} (each from the other)}, 
and there are only higher-order 
dependencies, if any. Then, {\color{black} $\C_\mathbbm{z}$ has the same 
block-diagonal 
	structure as $\C_\mathbbm{x}$}, so it is reasonable to select 
	$\R^{k\ell}={\bf 0}$ for $k\neq\ell$, although the sample covariances 
	$\wCz^{k\ell}$ are not exactly zero. In that 
case, \eqref{gradIVE} is simplified to {\color{black} (compare with 
\eqref{grad})}
\begin{equation}\label{gradIVEsimp}
\frac{\partial \mathcal{J}}{\partial {{\bf w}^k}^H}={\bf 
	a}^k-\frac{1}{N}\X^k\phi^k\bigl(\widehat{\bf s}^1,\dots,\widehat{\bf 
	s}^K\bigr)^T.
\end{equation} 

This observation is in agreement with the literature. The joint separation 
of 
correlated mixtures can be achieved using only second-order statistics  
\cite{li2009,lahat2016}. 
Uncorrelated mixtures arise, for instance, in the frequency-domain 
separation 
of convolutive mixtures. Here, the non-Gaussianity and higher-order moments 
are 
necessary for separating the mixtures \cite{kim2007,anderson2014}.

\subsection{Gradient algorithms for IVE}\label{sec:IVEalgorithms}
The constrained gradient can, similarly to \eqref{gradIVEsimp} 
and \eqref{gradw}, be computed with respect to $({\bf a}^k)^H$, which gives 
(we 
skip the detailed computation)
\begin{equation}\label{gradIVEsimpa}
\frac{\partial\mathcal{J}}{\partial{{\bf a}^k}^H}= 
{\bf w}^k-\frac{\lambda_{\bf 
		a}^k}{N}(\wCx^{kk})^{-1}\X^k\phi^k\bigl(\widehat{\bf 
	s}^1,\dots,\widehat{\bf 
	s}^K\bigr)^T,
\end{equation}
where $\lambda_{\bf a}^k=\bigl({{\bf a}^k}^H(\wCx^{kk})^{-1}{\bf 
	a}^k\bigr)^{-1}$. 
Now, the gradient optimization algorithms for IVE (considering only sets of 
uncorrelated instantaneous mixtures) can proceed in the same way as those 
for 
ICE with the following differences:
\begin{enumerate}
	\item In each iteration, ${\bf w}_k$ or ${\bf a}_k$ are updated by 
	adding a 
	step in the direction of the gradient \eqref{gradIVEsimp} or 
	\eqref{gradIVEsimpa}, respectively, for each $k=1,\dots,K$.
	\item The nonlinear functions $\phi^k$, $k=1,\dots,K$, depend on the 
	current 
	outputs of all $K$ iterative algorithms, which fact makes them mutually 
	dependent.
\end{enumerate}
We will refer to these algorithms as to OGIVE$_{\bf w}$ and OGIVE$_{\bf 
a}$, 
respectively; for illustration, OGIVE$_{\bf w}$ is described in 
Algorithm~\ref{algorithm4}.

\begin{algorithm}\label{algorithm4}
	\caption{OGIVE$_{\bf w}$: orthogonally constrained extraction of an 
		independent vector component from the set of mutually uncorrelated 
		mixtures 
		$\X^1,\dots,\X^K$}
	\SetAlgoLined
	\KwIn{$\X^k,{\bf w}_{\rm ini}^k$, $k=1,\dots,K$, $\mu$, ${\tt tol}$}
	\KwOut{${\bf a}^k,{\bf w}^k$, $k=1,\dots,K$}
	\ForEach{$k=1,\dots,K$}{
		$\widehat\C_\x^{kk}=\X^k(\X^k)^H/N$;\\
		${\bf w}^k={\bf w}_{\rm ini}^k$;}
	\Repeat{$\max\{\|\Delta^1\|,\dots,\|\Delta^K\|\}<{\tt tol}$}{
		\ForEach{$k=1,\dots,K$}{
			${\bf a}^k \leftarrow (({\bf w}^k)^H\widehat\C_\x^{kk}{\bf 
				w}^k)^{-1}\widehat\C_\x^{kk}{\bf w}^k$;\\
			$\widehat{\bf s}^k \leftarrow ({\bf w}^k)^H\X^k$;}
		\ForEach{$k=1,\dots,K$}{
			$\nu^k \leftarrow \widehat{\bf s}^k\phi^k(\widehat{\bf 
				s}^1,\dots,\widehat{\bf s}^K)^T/N$;\\
			$\Delta^k \leftarrow {\bf 
			a}^k-(\nu^k)^{-1}\X^k\phi^k(\widehat{\bf 
				s}^1,\dots,\widehat{\bf s}^K)^T/N$;\\
			${\bf w}^k \leftarrow {\bf w}^k + \mu\Delta^k$;}
	}
\end{algorithm}

{\color{black} Finally, we introduce a method referred to as OGIVE$_{\bf 
s}$, 
	where the idea presented in Section~\ref{sec:switchedopt} is applied 
	within 
	each mixture. The parameter in which the gradient optimization proceeds 
	is 
	selected based on the heuristic criterion \eqref{critoptim}. It is 
	important 
	that the selection can be different for each mixture. 
	
	In fact, this approach inherently assumes that the behavior of the 
	optimization 
	process within any mixture is similar as in case of ICE. The behavior 
	can be 
	different in case of IVE as the parallel extraction algorithms 
	influence each 
	other. The advantage of this feature is that the dependence can bring a 
	synergic effect: When most initial separating/mixing vectors lie in the 
	ROC of 
	the SOI, the convergence within the other mixtures can be enforced.     
	%As 
	%shown in Section~\ref{sec:switchedopt}, the effectiveness of this 
	%optimization 
	%depends on the initial SIR, which can be different for each mixture.

}

\newpage

\section{Simulations}\label{section:simulations}
\begin{figure*}
	\centering
	\includegraphics[width=\linewidth]{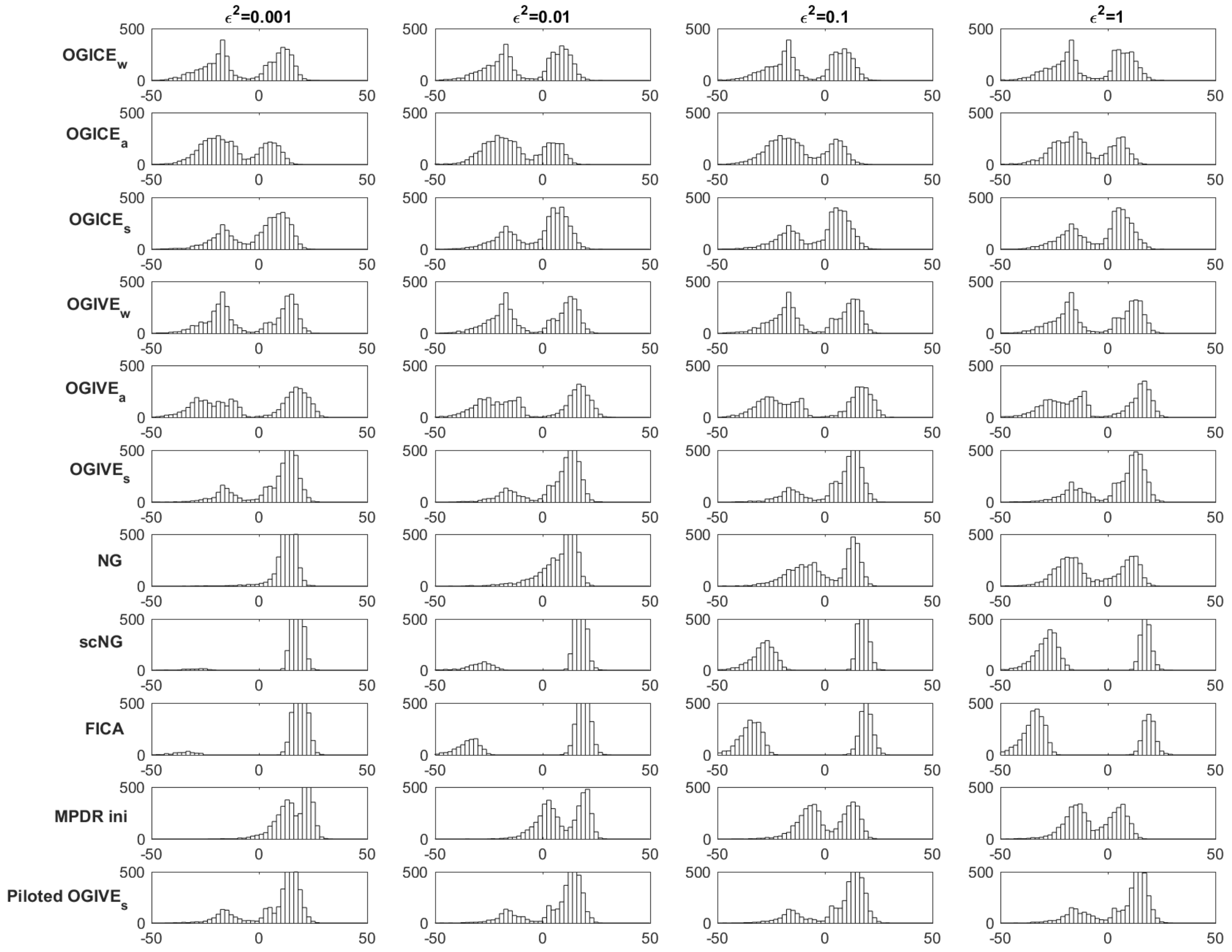}
	\caption{Histograms of the output SIR (in dB) achieved by the blind 
		algorithms in $1000$ trials ($4000$ extractions) when the 
		background is 
		Laplacean. 
		\label{fig:histograms}}
\end{figure*}

%The real-valued as well as the (circular) 
%complex-valued versions of the experiments are considered.
{\color{black} In simulations, we focus on the sensitivity of the proposed 
ICE 
	and IVE algorithms to the initialization and compare them with other 
	methods.}
In one simulated trial, $K$ instantaneous mixtures are generated with 
random 
mixing matrices $\A^k$, 
$k=1,\dots,K$. The SOIs for these mixtures are obtained as $K$ signals 
drawn 
independently from the Laplacean distribution and mixed 
by a random unitary matrix; hence, they are uncorrelated and dependent. The 
background is obtained by generating independent components 
$u_2^k,\dots,u_d^k$ from the Gaussian or Laplacean distribution, 
$k=1,\dots,K$; 
all the distributions are circular. 

{\color{black} For each mixture, the ${\rm SR}$ is selected randomly either 
	$-10$ 
	or $10$~dB. The mixing matrices are drawn from the uniform 
	distribution; the 
	real part in 
	$[1;2]$ and the imaginary part in $[0,1]$. This choice helps us keep 
	the 
	initial SIR approximately equal across all input channels, that is, 
	less 
	dependent on the mixing matrix while mostly dependent of the ${\rm 
	SR}$. }

The comparison involves One-Unit FastICA (FICA) \cite{bingham2000}, three 
variants of OGICE proposed in Section~\ref{sec:algorithms}, the Natural 
Gradient algorithm (NG) \cite{amari1996} and its scaled version (scNG) 
\cite{douglas2007},{\color{black} which is frequently used in audio 
separation 
	methods,} as well as three variants of OGIVE proposed in 
Section~\ref{sec:IVEalgorithms}. These algorithms are initialized by ${\bf 
	a}_{\rm ini}={\bf a} + {\bf e}_{\rm ini}$, where ${\bf a}$ is the true 
	mixing 
vector, and ${\bf e}_{\rm ini}$ is a random vector which is orthogonal to 
${\bf 
	a}$, and $\|{\bf e}_{\rm ini}\|^2=\epsilon^2$. The algorithms NG and 
	scNG 
are initialized by the de-mixing matrix whose first row is equal to 
\eqref{couplingw} where ${\bf a}={\bf a}_{\rm ini}$; the other rows are 
selected as in \eqref{demixing}, which means that the initial background 
subspace is orthogonal to the initial SOI estimate. 

Next, we also evaluate the SOI estimates obtained through \eqref{couplingw} 
for 
${\bf a}$ equal to the true mixing 
vector (MPDR\rule{2mm}{0.2mm}oracle) and for ${\bf a}={\bf a}_{\rm ini}$ 
(MPDR\rule{2mm}{0.2mm}ini). While the 
performance of the MPDR\rule{2mm}{0.2mm}oracle gives an upper bound, that 
of MPDR\rule{2mm}{0.2mm}ini 
corresponds to a ``do-nothing'' solution purely relying on the 
initialization. 
%An improvement is achieved by a blind algorithm when its performance is 
%higher 
%than that of MPDR~ini.

ICE/ICA methods are applied to each mixture separately, while IVE 
algorithms 
treat all $K$ mixtures jointly\footnote{\color{black} IVE can take 
advantage 
of 
	the dependence among SOIs from different mixtures while ICA/ICE cannot. 
	The 
	intention here is not to favor IVE prior to ICE; the goal is to 
	evaluate the 
	contribution due to the joint extraction.}.
The nonlinearity $\phi(\xi)=\overline{\tanh}(\xi)$ is selected in the 
variants 
of OGICE and NG. FICA is used with $(1+|\xi|^2)^{-1}$.
For IVE algorithms, the choice is
\begin{equation}\label{ivephi}
\phi^k(\xi^1,\dots,\xi^K)=\overline{\tanh}(\xi^k)/
\sqrt{\sum_{\ell=1}^K|\xi^\ell|^2}.
\end{equation}
The problem of choosing an appropriate nonlinearity for the given method 
would 
go 
beyond the scope of this paper; see, e.g., 
\cite{tichavsky2006,pham1997}.

For the sake of completeness, we also include a semi-blind variant of 
OGIVE$_{\bf s}$, which is modified in a way similar to that proposed in 
\cite{nesta2017}. Specifically, a ``pilot'' component $p$ is assumed to be 
available such that the SOIs within the $K$ mixtures are dependent on it 
(usually there are only higher-order dependencies; see \cite{kim2007}). 
OGIVE$_{\bf s}$ is modified only by adding the $K+1$th variable into 
\eqref{ivephi}, which is $\xi_{K+1}=p$. In simulations,  $p$ is a 
random mixture of the SOIs. This method will be referred 
to as ``Piloted OGIVE$_{\bf s}$''.

\begin{table}[]
	\centering
	\caption{Detailed settings of the compared methods}
	\label{table1}
	{\color{black}
		\begin{tabular}{l|ccc}
			\textbf{Algorithm(s)} & 
			\textbf{\begin{tabular}[c]{@{}c@{}}maximum\\ \# 
					iterations\end{tabular}} & 
			\multicolumn{1}{c}{\textbf{\begin{tabular}[c]{@{}c@{}}stopping\\
			 							threshold\end{tabular}}} & 
						\textbf{\begin{tabular}[c]{@{}c@{}}step\\ 
					length\end{tabular}} \\
			\hline
			OGICE variants     & 
			$5\times 
			10^3$                                                           
			     				&
			
			10$^{-3}$                                                       
			       				&
			0.1                                                            
			\\
			OGIVE variants     & 
			$4\times 
			10^3$                                                           
			&
			10$^{-3}$                                                       
			&
			0.1                                                            
			\\
			NG, scNG           & 
			$5\times 
			10^3$                                                          
			&
			10$^{-3}$                                                       
			&
			0.02                                                           
			\\
			FICA               & 
			$1\times 
			10^3$                                                           
			&
			10$^{-6}$                                                       
			&
			-                                                             
		\end{tabular}
	}
\end{table}
{\color{black}
	The detailed settings of the compared algorithms are shown in 
	Table~\ref{table1}; these values 
	were selected to ensure good performance of the methods.
}

\begin{figure*}
	\centering
	\includegraphics[width=\linewidth]{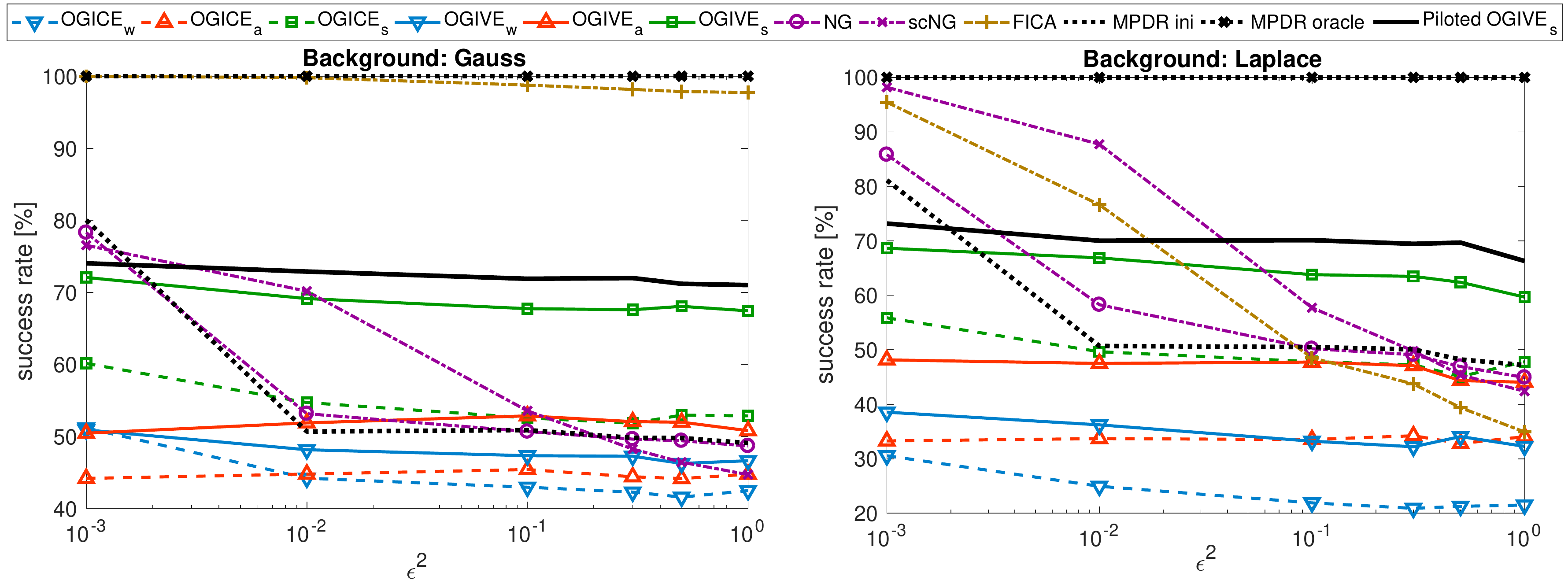}
	\caption{Success rates as functions of the initial mean squared error
		$\epsilon^2$ for $d=6$, $K=4$, and $N=1000$. The experiment was 
		repeated in 
		$1000$ trials ($1000\times K$ mixtures). The SOI is generated from 
		Laplace 
		components while the background is Gaussian or Laplacean.  
		\label{fig:successsigma}}
\end{figure*}

{\color{black}
	\subsection{Results}
	
	The algorithms were tested in $1000$ independent trials for $d=6$, 
	$K=4$, and 
	$N=1000$. Each extracted signal was assessed by the output SIR (the 
	ratio of 
	powers of the SOI and of the other signals within the extracted 
	signal). In the 
	experiment, the achieved SIRs range from $-50$ through $50$~dB 
	depending on 
	whether the SOI was extracted or a different source, and the SIR also 
	depends 
	on the qualities of the given algorithm. Complete results achieved by 
	the 
	methods, for the Laplacean background, are shown in 
	Fig.~\ref{fig:histograms} 
	as histograms.
	
	Since our primary focus is on the extraction of the SOI, the results 
	in Fig.~\ref{fig:successsigma} show the percentages of successful 
	extractions of the SOI (success rates) as functions of the initial 
	error 
	$\|{\bf e}_{\rm ini}\|^2=\epsilon^2$. Here, each extraction is classed 
	as {\em 
		successful} if the output SIR is greater than $0$~dB. We first 
		discuss these 
	results as follows.
	
	\begin{enumerate}
		
		\item {\bf MPDR:} MPDR~oracle achieves $100$\% success rate. The 
		success rate 
		of MPDR\rule{2mm}{0.2mm}ini is decaying with growing $\epsilon^2$ 
		as MPDR\rule{2mm}{0.2mm}ini yields SIR 
		corresponding to the extracted signal using the initial mixing 
		vector. Its 
		success rate approaches $50$\% as $\epsilon^2\rightarrow 1$, 
		which corresponds with the fact that ${\rm SR}=10$~dB in about 
		$50$\% of trials.
		The results of MPDR\rule{2mm}{0.2mm}oracle and of 
		MPDR\rule{2mm}{0.2mm}ini do not depend on the signals' 
		distributions.

		\item {\bf OGICE$_{{\bf a}/{\bf w}}$:} OGICE$_{\bf a}$ and 
		OGICE$_{\bf w}$ 
		achieve success rate 20-50\% almost independently of $\epsilon^2$. 
		This 
		corresponds with the fact that maximum 50\% 
		of trials are advantageous for each method (${\rm SR}=-10$dB for 
		OGICE$_{\bf 
			a}$ and ${\rm SR}=+10$dB for OGICE$_{\bf a}$). The success 
			rates are mostly 
		lower than 50\%, which shows that also the mixing matrix has an 
		influence of 
		the ROC of the SOI, so, for example, ${\rm SR}=+10$dB does not 
		always guarantee 
		that OGICE$_{\bf w}$ converges to the SOI from almost any initial 
		value\footnote{\color{black} It can be verified, by repeating this 
			experiment, 
			that with 
			${\rm SR}=\pm 15$dB, OGICE$_{\bf a}/{\bf w}$ achieve almost 
			50\% 
			success rate. It means that with a higher range of ${\rm SR}$ 
			the influence of 
			the mixing matrix (generated in the same vein as in this 
			experiment) is lower.}. For the Laplacean background, the 
			success rates 
		are lower than for the Gaussian background because there is a 
		higher 
		probability that the methods are attracted by a different 
		non-Gaussian source.
		
		\item {\bf OGIVE$_{{\bf a}/{\bf w}}$:} The IVE counterparts of  
		OGICE$_{\bf a}$ 
		and OGICE$_{\bf w}$ perform similarly but their success rates are 
		always  
		closer to $50\%$. This is caused by the joint extraction. 
		OGIVE$_{\bf a}$ and 
		OGIVE$_{\bf w}$ take advantage of the dependence between the SOIs 
		in $K$ 
		mixtures, which improves the overall convergence.
		
		\item {\bf Switched optimization:} In terms of the success rate, 
		OGICE$_{\bf 
			s}$ and OGIVE$_{\bf s}$ achieve significant improvements 
			compared to 
		OGICE$_{{\bf a}/{\bf w}}$ and OGIVE$_{{\bf a}/{\bf w}}$, 
		respectively.  This 
		points to the effectiveness of the decision rule based on the 
		criterion 
		\eqref{critoptim} and to the synergy of the joint separation in 
		case of 
		OGIVE$_{\bf s}$. It is worth pointing to the fact that the 
		performance of 
		OGICE$_{\bf s}$ and OGIVE$_{\bf s}$ is slightly decreasing with 
		growing 
		$\epsilon^2$. This is because for higher $\epsilon^2$, the initial 
		mixing 
		vector is more distant from the true value, so the criterion 
		\eqref{critoptim} is less reliable for the selection of the 
		optimization 
		parameter. To avoid this drawback, a better decision rule or 
		partial knowledge 
		of ${\rm SR}$) is needed.
		
		\item {\bf NG and scNG:} The success rate of these ICA algorithms 
		significantly 
		depends on the initial error and also on the distribution of the 
		background. 
		The success rate is superior for very small values of $\epsilon^2$, 
		but it 
		rapidly drops with growing $\epsilon^2$ (scNG appears to be less 
		sensitive than 
		NG). The results also show that very good convergence is achieved 
		for 
		$\epsilon^2<10^{-2}$ when the background is Laplacean. This points 
		to the fact 
		that ICA algorithms can take advantage of the non-Gaussianity of 
		background 
		signals.
		
		\item {\bf FICA} shows excellent results when the background is 
		Gaussian. Since it is a fixed-point algorithm, it has good ability 
		to 
		avoid shallow extremes of the contrast function that correspond to 
		Gaussian components. Therefore, its global convergence is very good 
		when 
		there are no other non-Gaussian components than the SOI. For the 
		same 
		reason, the success rate of FICA is dropping down with growing 
		$\epsilon^2$  
		when the background is Laplacean.

		\item {\bf Piloted OGIVE$_{\bf s}$}: This algorithm gives a
		higher success rate than OGIVE$_{\bf s}$ as it exploits the pilot 
		dependent 
		component to keep converging to the SOI. Its performance is 
		slightly decreasing 
		when $\epsilon^2$ grows due to the shortage of the criterion 
		\eqref{critoptim}, 
		as mentioned in Item 4 above.
		
	\end{enumerate}
	
	Now we go back to the histograms of the output SIR (in dB) shown in 
	Fig.~\ref{fig:histograms}. Typically, the values are concentrated 
	around two 
	peaks with central value $\approx\pm 20$~dB. When more values are 
	concentrated 
	around the positive SIR for any $\epsilon^2$, the given method shows 
	good robustness against the 
	initialization error as it mostly tends to keep converging to the SOI. 
	From 
	this perspective, OGICE$_{\bf s}$, OGIVE$_{\bf s}$ and Piloted 
	OGIVE$_{\bf s}$ 
	show the best results, which was already discussed above.
	
	The variance of the SIRs around the peaks reflect the ability of the 
	method to 
	avoid local extremes of the contrast function and/or its ability to 
	converge 
	before the maximum number of iterations is reached. In this respect, 
	scNG and 
	FICA achieve superior results as they rarely yield an 
	output SIRs within the range $[-10,10]$~dB. It is worth pointing out 
	that scNG 
	and FICA could be interpreted as gradient methods using adaptive step 
	lengths, 
	as discussed in Section~\ref{sec:fastica} for the case of FICA. The 
	other 
	compared algorithms, the variants of OGICE and of OGIVE and NG, utilize 
	constant step lengths. Their histograms are less concentrated around 
	the main 
	peaks, which means that they often stack in a local extreme or converge 
	too 
	slowly to achieve the desired extreme of the contrast function.
}

{\color{black}
	\section{Conclusions}
	We have revised the problem of blind source extraction of an 
	independent target 
	source from background signals. The maximum likelihood approach where 
	the 
	mixing model is parameterized for the extraction of one source and 
	where the 
	background signals are modeled as a Gaussian mixture was introduced as 
	Independent Component Extraction (ICE). Similarly, Independent Vector 
	Extraction (IVE) was introduced for the joint extraction problem. 
	
	Several variants of gradient algorithms have been derived. Our 
	attention has 
	been focused on their region of convergence (ROC) related to the source 
	of 
	interest (SOI). It was shown that the size of the ROC is not only 
	algorithm-dependent but it also strongly depends on the ratio of scales 
	of the 
	sources within the mixture. In particular, we have shown that the size 
	of the 
	ROC can be influenced through the selection of optimization parameters. 
	This was corroborated by simulations where the 
	methods endowed with the automatic selection of optimization parameters 
	achieved a high rate of 
	successful extractions of the SOI, almost independent of the 
	initialization.
	
	The simulation study also confirmed that the joint extraction through 
	IVE 
	brings advantageous features compared ICE. In particular, the IVE 
	methods with 
	the automatic selection show synergistic convergence, by which we mean 
	that 
	simultaneous converge to the SOIs within several mixtures help to 
	enforce the 
	convergence also in the other mixtures.
	
	Next, the gradient algorithms derived based on ICE have been compared 
	with 
	the Natural Gradient-based methods for ICA and with One-Unit FastICA. 
	Close 
	connections between the methods have been revealed, which sheds light 
	on the 
	relation between ICE and the well known blind source 
	separation/extraction 
	(BSS/BSE) methods. In particular, the importance of the orthogonal 
	constraint 
	(the orthogonality of the subspaces spanned by the SOI and the 
	background 
	signals) and the Gaussian modeling of the background signals in ICE 
	methods was 
	shown. Therefore, future works should be focused on these aspects in 
	order to 
	improve  overall properties of ICE/IVE algorithms in non-Gaussian 
	background 
	and also in underdetermined scenarios.
}

\section*{Appendix A: Proof of \eqref{couplingw} and \eqref{couplinga}}

Since $\y={\bf Qz}={\bf QBx}$, we can introduce the projection operator 
$\boldsymbol{\Pi}_\y={\bf QB}$, which is equal to
\begin{equation}\label{projectory}
\boldsymbol{\Pi}_\y={\bf I}_d-{\bf a}{\bf w}^H.
\end{equation}
According to \eqref{orthconst}, the OC can be written as
\begin{equation}
\widehat{\bf Z}^H\,\widehat{{\bf s}}/N={\bf B}\widehat\C_\x{\bf w}={\bf 0}.
\end{equation}
By multiplying the latter equation from the left by ${\bf Q}$ and using 
\eqref{projectory}, we arrive at
\begin{align}
({\bf I}_d-{\bf a}{\bf w}^H)\widehat\C_\x{\bf w}&={\bf 0},\\
{\bf w}-\widehat\C_\x^{-1}{\bf a}\,({\bf w}^H\widehat\C_\x{\bf w})&={\bf 
	0}.\label{aux1}
\end{align}
By multiplying \eqref{aux1} from the left by ${\bf a}^H$ it follows
\begin{equation}
{\bf a}^H{\bf w}-{\bf a}^H\widehat\C_\x^{-1}{\bf a}\,{\bf 
w}^H\widehat\C_\x{\bf 
	w}={\bf 0},
\end{equation}
and since ${\bf a}^H{\bf w}=1$, it holds that
\begin{equation}
{\bf a}^H\widehat\C_\x^{-1}{\bf a}=({\bf w}^H\widehat\C_\x{\bf 
	w})^{-1}\label{aux2}.
\end{equation}
By putting \eqref{aux1} and \eqref{aux2} together, \eqref{couplingw} and 
\eqref{couplinga} follow.\hfill\rule{1.2ex}{1.2ex}

\section*{Appendix B: Proof of \eqref{optimumMPDR}}
Assume that ${\bf a}$ is equal to its true value in 
\eqref{modelsinglesource} 
(hence $\A=\A_{\rm ICE}$ and $\W=\W_{\rm ICE}$), 
and recall that, in the determined ICE model, ${\bf y}={\bf Qz}$ holds.
Then, ${\bf w}_{\rm MPDR}^H\x=s+{\bf w}_{\rm MPDR}^H\y$. We should show 
that 
${\bf a}^H\Cx^{-1}\y={\bf 0}$. It holds that
\begin{align}
{\bf a}^H\Cx^{-1}\y&={\bf a}^H\Cx^{-1}{\bf Qz}\\
&={\bf a}^H\left(\A\C_{\bf v}\A^H\right)^{-1}{\bf Qz}\\
&={\bf a}^H\W^H\C_{\bf v}^{-1}\W{\bf Qz},
\end{align}
where $\C_{\bf v}={\rm E}[{\bf v}{\bf v}^H]$. Next, it holds that 
${\bf a}^H\W^H=[1,\,{\bf 0}^H]$, and $\W{\bf Q}=[{\bf 0},\,{\bf 
	I}_{d-1}]^H$. By taking into account the block-diagonal 
structure of $\C_{\bf v}$, i.e.,
\begin{equation}
\C_{\bf v}=\begin{pmatrix}
\sigma^2_s & {\bf 0}^H\\
{\bf 0} & \C_\z
\end{pmatrix},
\end{equation}
where $\sigma^2_s$ denotes the variance of $s$, \eqref{optimumMPDR} 
follows.\hfill\rule{1.2ex}{1.2ex}

\section*{Appendix C: Computation of \eqref{grad1}}
The following identities hold under the constraint \eqref{couplinga}.
\begin{align}
{\bf g}&={\bf E}\,{\bf a}=\frac{{\bf E}\wCx{\bf w}}{{\bf w}^H\wCx{\bf w}},\\
\gamma&={\bf e}_1^H{\bf a}=\frac{{\bf e}_1^H\wCx{\bf w}}{{\bf w}^H\wCx{\bf 
		w}},\\
1-\h^H\g&=\overline{\beta}\gamma=\overline{\beta}\frac{{\bf e}_1^H\wCx{\bf 
		w}}{{\bf w}^H\wCx{\bf w}}.
\end{align}
To derive \eqref{grad1}, we proceed by computing the derivatives of 
the three terms in \eqref{contrastcon}. First, using \eqref{score}, it 
follows 
that
\begin{equation}
\frac{\partial}{\partial {\bf w}^H}\log f({\bf w}^H\x) = -\phi({\bf 
w}^H\x)\x,
\end{equation}
so that
\begin{equation}\label{ac1}
\frac{1}{N}\sum_{n=1}^N-\phi\bigl({\bf 
	w}^H\x(n)\bigr)\x(n)=-\frac{1}{N}\X\phi({\bf 
	w}^H\X)^T, 
\end{equation}
where $\phi(\cdot)$ is applied element-wise in case of the vector argument.

Let $\x$ be partitioned as $\x=[x_1;\x_2]$. Under the constraint 
\eqref{couplinga} 
and using ${\bf B}=[\g, \,-\gamma{\bf I}_{d-1}]$, the second term in 
\eqref{contrastcon} (where the argument $n$ is omitted) can be re-written as
\begin{multline}
\x^H{\bf B}^H{\bf R}{\bf B}\x=({\bf w}^H\wCx{\bf w})^{-1} \times \\ 
\bigl(|x_1|^2{\bf 
	w}^H\wCx{\bf E}^H{\bf R}{\bf E}\wCx{\bf w} - \overline{x_1}{\bf 
	w}^H\wCx{\bf 
	E}^H{\bf e}_1^H\wCx{\bf w}{\bf R}\x_2-\\ x_1{\bf w}^H\wCx{\bf 
	e}_1\x_2^H{\bf 
	R}{\bf E}\wCx{\bf w} + {\bf w}^H\wCx{\bf e}_1\x_2^H{\bf R}{\bf 
	e}_1^H\wCx{\bf 
	w}\x_2\bigr).
\end{multline}
By taking the derivative under the OG and after some rearrangements,
\begin{multline}\label{ac3}
\frac{\partial}{\partial {\bf w}^H}\x^H{\bf B}^H{\bf R}{\bf B}\x= -2{\bf 
	a}\,\x^H{\bf B}^H{\bf R}{\bf B}\x+\\({\bf w}^H\wCx{\bf w})^{-1}\times
\bigl(\overline{x_1}\wCx{\bf E}^H{\bf R}{\bf B}\x-\x_2^H{\bf R}{\bf 
	B}\x\wCx{\bf e}_1\bigr).
\end{multline}
By considering the averages of the above terms over $N$ samples, we arrive 
at 
the following chain of identities:
\begin{align}
\frac{1}{N}\sum_{n=1}^N \x(n)^H{\bf B}^H{\bf R}{\bf 
	B}\x(n)&=\frac{1}{N}\sum_{n=1}^N {\tt tr}({\bf B}^H{\bf R}{\bf 
	B}\x(n)^H\x(n))\nonumber\\
&={\tt tr}\bigl({\bf R}{\bf B}\wCx{\bf B}^H\bigr) ={\tt tr}\bigl({\bf 
	R}\wCz\bigr).\label{ac4}
\end{align}
Next,
\begin{align}
\frac{1}{N}\sum_{n=1}^N \overline{x_1}(n)\wCx{\bf E}^H&{\bf R}{\bf B}\x(n)=
\wCx{\bf E}^H{\bf R}{\bf B}\X\X^H{\bf e}_1/N\nonumber \\
=& \wCx{\bf E}^H{\bf R}\widehat{\bf Z}\X^H{\bf e}_1/N\nonumber\\
=& \wCx{\bf E}^H{\bf R}\widehat{\bf Z}\begin{pmatrix}
\widehat{\bf s}^H& \widehat{\bf Z}^H
\end{pmatrix}\A_{\rm ICE}^H{\bf e}_1/N\nonumber\\
=& \wCx{\bf E}^H{\bf R}\begin{pmatrix}
{\bf 0}^H& \wCz
\end{pmatrix}\A_{\rm ICE}^H{\bf e}_1\nonumber\\
=& \wCx{\bf E}^H{\bf R}\wCz\h,\label{ac5}
\end{align}
where we used \eqref{demixedsignals} and \eqref{orthconst}. 
The last identity is
\begin{equation}\label{ac6}
\frac{1}{N}\sum_{n=1}^N \x_2^H{\bf R}{\bf B}\x\wCx{\bf e}_1 =
{\tt tr}\bigl({\bf R}{\bf B}\wCx{\bf E}^H\bigr)\wCx{\bf e}_1.
\end{equation}

The derivative of the third term in \eqref{contrastcon} reads
\begin{multline}\label{ac2}
(d-2)\frac{\partial}{\partial {\bf w}^H}\log |\gamma|^2 = \\
(d-2)\frac{\partial}{\partial {\bf w}^H}( \log|{\bf w}^H\wCx{\bf e}_1|^{2} 
- 
\log|{\bf w}^H\wCx{\bf w}|^{2})=\\
(d-2)\frac{\partial}{\partial {\bf w}^H}( \log {\bf w}^H\wCx{\bf e}_1 -
2\log {\bf w}^H\wCx{\bf w})=\\
(d-2)\left(\frac{\wCx{\bf e}_1}{{\bf w}^H\wCx{\bf e}_1}-2\frac{\wCx{\bf 
		w}}{{\bf w}^H\wCx{\bf w}}\right)=\\
(d-2)\left(\overline{\gamma}^{-1}\frac{\wCx{\bf e}_1}{{\bf w}^H\wCx{\bf 
		w}}-2{\bf a}\right).
\end{multline}

Now, \eqref{grad1} is obtained by putting together \eqref{ac1}, 
\eqref{ac3}, 
and 
\eqref{ac2} using the identities \eqref{ac4}, \eqref{ac5}, and 
\eqref{ac6}.
\hfill\rule{1.2ex}{1.2ex}

%	\bibliographystyle{IEEEtran}
% Generated by IEEEtran.bst, version: 1.14 (2015/08/26)

%	\bibliography{IEEEfull,conferences,ISI}

\begin{thebibliography}{10}
\providecommand{\url}[1]{#1}
\csname url@samestyle\endcsname
\providecommand{\newblock}{\relax}
\providecommand{\bibinfo}[2]{#2}
\providecommand{\BIBentrySTDinterwordspacing}{\spaceskip=0pt\relax}
\providecommand{\BIBentryALTinterwordstretchfactor}{4}
\providecommand{\BIBentryALTinterwordspacing}{\spaceskip=\fontdimen2\font plus
\BIBentryALTinterwordstretchfactor\fontdimen3\font minus
  \fontdimen4\font\relax}
\providecommand{\BIBforeignlanguage}[2]{{%
\expandafter\ifx\csname l@#1\endcsname\relax
\typeout{** WARNING: IEEEtran.bst: No hyphenation pattern has been}%
\typeout{** loaded for the language `#1'. Using the pattern for}%
\typeout{** the default language instead.}%
\else
\language=\csname l@#1\endcsname
\fi
#2}}
\providecommand{\BIBdecl}{\relax}
\BIBdecl

\bibitem{lee1998}
T.-W. Lee, \emph{Independent Component Analysis - Theory and
  Applications}.\hskip 1em plus 0.5em minus 0.4em\relax Kluwer Academic
  Publishers, 1998.

\bibitem{hyvarinen2001}
A.~Hyv\"{a}rinen, J.~Karhunen, and E.~Oja, \emph{Independent Component
  Analysis}.\hskip 1em plus 0.5em minus 0.4em\relax John Wiley \& Sons, 2001.

\bibitem{cichocki2002}
A.~Cichocki and S.~Amari, \emph{Adaptive Blind Signal and Image
  Processing}.\hskip 1em plus 0.5em minus 0.4em\relax John Wiley \& Sons, 2002.

\bibitem{comon2010handbook}
P.~Comon and C.~Jutten, \emph{Handbook of Blind Source Separation: Independent
  Component Analysis and Applications}, ser. Independent Component Analysis and
  Applications Series.\hskip 1em plus 0.5em minus 0.4em\relax Elsevier Science,
  2010.

\bibitem{smaragdis1998}
P.~Smaragdis, ``Blind separation of convolved mixtures in the frequency
  domain,'' \emph{Neurocomputing}, vol.~22, pp. 21--34, 1998.

\bibitem{li2009}
Y.~O. Li, T.~Adal\i, W.~Wang, and V.~D. Calhoun, ``Joint blind source
  separation by multiset canonical correlation analysis,'' \emph{IEEE
  Transactions on Signal Processing}, vol.~57, no.~10, pp. 3918--3929, Oct
  2009.

\bibitem{lahat2014}
D.~Lahat and C.~Jutten, ``Joint blind source separation of multidimensional
  components: Model and algorithm,'' in \emph{2014 22nd European Signal
  Processing Conference (EUSIPCO)}, Sept 2014, pp. 1417--1421.

\bibitem{li2011}
X.-L. Li, T.~Adal\i, and M.~Anderson, ``Joint blind source separation by
  generalized joint diagonalization of cumulant matrices,'' \emph{Signal
  Processing}, vol.~91, no.~10, pp. 2314 -- 2322, 2011.

\bibitem{kim2007}
T.~Kim, H.~T. Attias, S.-Y. Lee, and T.-W. Lee, ``Blind source separation
  exploiting higher-order frequency dependencies,'' \emph{{IEEE} Transactions
  on Audio, Speech, and Language Processing}, pp. 70--79, Jan. 2007.

\bibitem{sawada2004icakyoto}
H.~Sawada, R.~Mukai, S.~Araki, and S.~Makino, ``Solving the permutation and the
  circularity problem of frequency-domain blind source separation,'' in
  \emph{Proc. of the 18th International Congress on Acoustics (ICA 2004)}, Apr.
  2004, pp. I--89--I--92.

\bibitem{koldovsky2013}
Z.~Koldovsk\'y, J.~M\'alek, P.~Tichavsk\'y, and F.~Nesta, ``Semi-blind noise
  extraction using partially known position of the target source,'' \emph{IEEE
  Transactions on Audio, Speech, and Language Processing}, vol.~21, no.~10, pp.
  2029--2041, Oct 2013.

\bibitem{sawada2005icassp}
H.~Sawada, S.~Araki, R.~Mukai, and S.~Makino, ``Blind extraction of a dominant
  source signal from mixtures of many sources,'' in \emph{Proceedings of {IEEE}
  International Conference on Audio, Speech and Signal Processing}, vol. III,
  Mar. 2005, pp. 61--64.

\bibitem{javidi2010}
S.~Javidi, D.~P. Mandic, and A.~Cichocki, ``Complex blind source extraction
  from noisy mixtures using second-order statistics,'' \emph{IEEE Transactions
  on Circuits and Systems I: Regular Papers}, vol.~57, no.~7, pp. 1404--1416,
  July 2010.

\bibitem{amari1998b}
S.~Amari and A.~Cichocki, ``Adaptive blind signal processing-neural network
  approaches,'' \emph{Proceedings of the IEEE}, vol.~86, no.~10, pp.
  2026--2048, Oct 1998.

\bibitem{cruces2004}
S.~A. Cruces-Alvarez, A.~Cichocki, and S.~Amari, ``From blind signal extraction
  to blind instantaneous signal separation: criteria, algorithms, and
  stability,'' \emph{IEEE Transactions on Neural Networks}, vol.~15, no.~4, pp.
  859--873, July 2004.

\bibitem{huber1985}
P.~J. Huber, ``Projection pursuit,'' \emph{Ann. Statist.}, vol.~13, no.~2, pp.
  435--475, June 1985.

\bibitem{girolami1997}
M.~Girolami, ``\BIBforeignlanguage{English}{Extraction of independent signal
  sources using a deflationary exploratory projection pursuit network with
  lateral inhibition},'' \emph{\BIBforeignlanguage{English}{IEE Proceedings -
  Vision, Image and Signal Processing}}, vol. 144, pp. 299--306(7), October
  1997.

\bibitem{shalvi1993}
O.~Shalvi and E.~Weinstein, ``Super-exponential methods for blind
  deconvolution,'' \emph{IEEE Transactions on Information Theory}, vol.~39,
  no.~2, pp. 504--519, Mar 1993.

\bibitem{shynk1996}
J.~J. Shynk and R.~P. Gooch, ``The constant modulus array for cochannel signal
  copy and direction finding,'' \emph{IEEE Transactions on Signal Processing},
  vol.~44, no.~3, pp. 652--660, Mar 1996.

\bibitem{moreau1996}
E.~Moreau and O.~Macchi, ``High-order contrasts for self-adaptive source
  separation,'' \emph{International Journal of Adaptive Control and Signal
  Processing}, vol.~10, no.~1, pp. 19--46, 1996.

\bibitem{cover}
T.~Cover and J.~Thomas, \emph{Elements of Information Theory}.\hskip 1em plus
  0.5em minus 0.4em\relax Wiley, 2006.

\bibitem{cardoso1998}
J.~F. Cardoso, ``Blind signal separation: statistical principles,''
  \emph{Proceedings of the {IEEE}}, vol.~86, no.~10, pp. 2009--2025, Oct 1998.

\bibitem{delfosse1995}
N.~Delfosse and P.~Loubaton, ``Adaptive blind separation of independent
  sources: A deflation approach,'' \emph{Signal Processing}, vol.~45, no.~1,
  pp. 59 -- 83, 1995.

\bibitem{erdogan2009}
A.~T. Erdogan, ``On the convergence of ica algorithms with symmetric
  orthogonalization,'' \emph{IEEE Transactions on Signal Processing}, vol.~57,
  no.~6, pp. 2209--2221, June 2009.

\bibitem{hyvarinen1997}
A.~Hyv\"{a}rinen and E.~Oja, ``A fast fixed-point algorithm for independent
  component analysis,'' \emph{Neural Computation}, vol.~9, no.~7, pp.
  1483--1492, July 1997.

\bibitem{hyvarinen1999}
A.~Hyv\"{a}rinen, ``Fast and robust fixed-point algorithm for independent
  component analysis,'' \emph{{IEEE} Transactions on Neural Networks}, vol.~10,
  no.~3, pp. 626--634, 1999.

\bibitem{wei2015}
T.~Wei, ``A convergence and asymptotic analysis of the generalized symmetric
  fastica algorithm,'' \emph{IEEE Transactions on Signal Processing}, vol.~63,
  no.~24, pp. 6445--6458, Dec 2015.

\bibitem{tichavsky2006}
P.~Tichavsk\'y, Z.~Koldovsk\'y, and E.~Oja, ``Performance analysis of the
  {FastICA} algorithm and {Cram\'er-Rao} bounds for linear independent
  component analysis,'' \emph{{IEEE} Transactions on Signal Processing},
  vol.~54, no.~4, pp. 1189--1203, April 2006.

\bibitem{hyvarinen1997b}
A.~Hyv\"{a}rinen, ``One-unit contrast functions for independent component
  analysis: a statistical analysis,'' in \emph{Neural Networks for Signal
  Processing VII. Proceedings of the 1997 IEEE Signal Processing Society
  Workshop}, Sep 1997, pp. 388--397.

\bibitem{cardoso1994}
J.-F. Cardoso, ``On the performance of orthogonal source separation
  algorithms,'' in \emph{Proceedings of European Signal Processing Conference},
  Sep. 1994, pp. 776--779.

\bibitem{koldovsky2006}
Z.~Koldovsk\'y, P.~Tichavsk\'y, and E.~Oja, ``Efficient variant of algorithm
  {FastICA} for independent component analysis attaining the {Cram\'er-Rao}
  lower bound,'' \emph{{IEEE} Transactions on Neural Networks}, vol.~17, no.~5,
  pp. 1265--1277, Sept 2006.

\bibitem{pham2006}
D.-T.~A. Pham, ``Blind partial separation of instantaneous mixtures of
  sources,'' in \emph{Proceedings of International Conference on Independent
  Component Analysis and Signal Separation}.\hskip 1em plus 0.5em minus
  0.4em\relax Springer Berlin Heidelberg, 2006, pp. 868--875.

\bibitem{koldovsky2017b}
Z.~Koldovsk\'{y}, P.~Tichavsk\'y, and V.~Kautsk\'y, ``Orthogonally constrained
  independent component extraction: Blind {MPDR} beamforming,'' in
  \emph{Proceedings of European Signal Processing Conference}, Sep. 2017, pp.
  1195--1199.

\bibitem{cardoso2017}
J.~F. Cardoso, ``On extracting the cosmic microwave background from
  multi-channel measurements,'' in \emph{Proceedings of International
  Conference on Latent Variable Analysis and Signal Separation}, Feb 2017, pp.
  403--413.

\bibitem{cardoso2008}
J.~F. Cardoso, M.~L. Jeune, J.~Delabrouille, M.~Betoule, and G.~Patanchon,
  ``Component separation with flexible models - application to multichannel
  astrophysical observations,'' \emph{IEEE Journal of Selected Topics in Signal
  Processing}, vol.~2, no.~5, pp. 735--746, Oct 2008.

\bibitem{lee2007iva}
I.~Lee, T.~Kim, and T.-W. Lee, ``Independent vector analysis for convolutive
  blind speech separation,'' in \emph{Blind speech separation}.\hskip 1em plus
  0.5em minus 0.4em\relax Springer, 2007, pp. 169--192.

\bibitem{adali2014}
T.~Adal{\i}, M.~Anderson, and G.~S. Fu, ``Diversity in independent component
  and vector analyses: Identifiability, algorithms, and applications in medical
  imaging,'' \emph{{IEEE} Signal Processing Magazine}, vol.~31, no.~3, pp.
  18--33, May 2014.

\bibitem{vantrees2002}
H.~L. Van~Trees, \emph{Optimum Array Processing: Part IV of Detection,
  Estimation, and Modulation Theory}.\hskip 1em plus 0.5em minus 0.4em\relax
  John Wiley \& Sons, Inc., 2002.

\bibitem{cardoso1998b}
J.~F. Cardoso, ``Multidimensional independent component analysis,'' in
  \emph{Proceedings of {IEEE} International Conference on Audio, Speech and
  Signal Processing}, vol.~4, May 1998, pp. 1941--1944 vol.4.

\bibitem{hyvarinen2006}
A.~Hyv\"{a}rinen and U.~K\"{o}ster, ``{FastISA}: A fast fixed-point algorithm
  for independent subspace analysis,'' in \emph{14th European symposium on
  artificial neural networks, (ESANN 2006)}, 2006.

\bibitem{matsuoka2001}
K.~Matsuoka and S.~Nakashima, ``Minimal distortion principle for blind source
  separation,'' in \emph{Proceedings of International Conference on Independent
  Component Analysis and Signal Separation}, Dec. 2001, pp. 722--727.

\bibitem{duong2010under}
N.~Duong, E.~Vincent, and R.~Gribonval, ``Under-determined reverberant audio
  source separation using a full-rank spatial covariance model,'' \emph{{IEEE}
  Transactions on Audio, Speech, and Language Processing}, vol.~18, no.~7, pp.
  1830--1840, 2010.

\bibitem{koldovsky2017}
Z.~Koldovsk\'{y} and F.~Nesta, ``Performance analysis of source image
  estimators in blind source separation,'' \emph{{IEEE} Transactions on Signal
  Processing}, vol.~65, no.~16, pp. 4166--4176, Aug. 2017.

\bibitem{sawada2004sap}
H.~Sawada, R.~Mukai, S.~Araki, and S.~Makino, ``A robust and precise method for
  solving the permutation problem of frequency-domain blind source
  separation,'' \emph{{IEEE} Transactions on Speech and Audio Processing},
  vol.~12, no.~5, pp. 530--538, Sep. 2004.

\bibitem{pham1997}
D.~T. Pham and P.~Garat, ``Blind separation of mixture of independent sources
  through a quasi-maximum likelihood approach,'' \emph{{IEEE} Transactions on
  Signal Processing}, vol.~45, no.~7, pp. 1712--1725, Jul 1997.

\bibitem{brandwood1983}
D.~H. Brandwood, ``A complex gradient operator and its application in adaptive
  array theory,'' \emph{Communications, Radar and Signal Processing, IEE
  Proceedings F}, vol. 130, no.~1, pp. 11--16, February 1983.

\bibitem{li2008}
H.~Li and T.~Adal{\i}, ``Complex-valued adaptive signal processing using
  nonlinear functions,'' \emph{EURASIP Journal on Advances in Signal
  Processing}, vol. 2008, Feb 2008, {Article ID 765615}.

\bibitem{bell1995}
A.~Bell and T.~Sejnowski, ``An information-maximization approach to blind
  separation and blind deconvolution,'' \emph{Neural Computation}, vol.~7,
  no.~6, pp. 1129--1159, 1995.

\bibitem{amari1996}
S.~Amari, A.~Cichocki, and H.~H. Yang, ``A new learning algorithm for blind
  signal separation,'' in \emph{Proceedings of Neural Information Processing
  Systems}, 1996, pp. 757--763.

\bibitem{laheld1996}
J.~F. Cardoso and B.~H. Laheld, ``Equivariant adaptive source separation,''
  \emph{{IEEE} Transactions on Signal Processing}, vol.~44, no.~12, pp.
  3017--3030, Dec 1996.

\bibitem{lahat2016}
D.~Lahat and C.~Jutten, ``Joint independent subspace analysis using
  second-order statistics,'' \emph{IEEE Transactions on Signal Processing},
  vol.~64, no.~18, pp. 4891--4904, Sept 2016.

\bibitem{anderson2014}
M.~Anderson, G.~S. Fu, R.~Phlypo, and T.~Adal\i, ``Independent vector analysis:
  Identification conditions and performance bounds,'' \emph{IEEE Transactions
  on Signal Processing}, vol.~62, no.~17, pp. 4399--4410, Sept 2014.

\bibitem{bingham2000}
E.~Bingham and A.~Hyv\"{a}rinen, ``A fast fixed-point algorithm for independent
  component analysis of complex valued signals,'' \emph{International Journal
  of Neural Systems}, vol.~10, no.~1, pp. 1--8, Feb. 2000.

\bibitem{douglas2007}
S.~C. Douglas and M.~Gupta, ``Scaled natural gradient algorithms for
  instantaneous and convolutive blind source separation,'' in \emph{2007 IEEE
  International Conference on Acoustics, Speech and Signal Processing - ICASSP
  '07}, vol.~2, April 2007, pp. II--637--II--640.

\bibitem{nesta2017}
F.~Nesta and Z.~Koldovsk\'y, ``Supervised independent vector analysis through
  pilot dependent components,'' in \emph{2017 IEEE International Conference on
  Acoustics, Speech and Signal Processing (ICASSP)}, March 2017, pp. 536--540.

\end{thebibliography}

\end{document}